\documentclass{article}

\usepackage{acra}

\usepackage{subcaption}
\usepackage{siunitx}
\usepackage{booktabs}
\usepackage{graphicx}
\usepackage{csquotes}
\usepackage{soul}
\usepackage{algpseudocode}
\usepackage{amsmath,amssymb}
\usepackage{algorithm2e}
\usepackage{float}
\usepackage{ragged2e}
\usepackage{adjustbox}
\usepackage{cite}

\usepackage{url}
\usepackage{multirow}
\usepackage{array}
\usepackage{multicol}

\newtheorem{Def}{Definition}[section]

\newcounter{example}[section]
\newenvironment{example}[1][]{\refstepcounter{example}\par\medskip
   \noindent \textbf{Example~\theexample. #1} \rmfamily}{\medskip}

\begin{document}


\onecolumn
\begin{center}
\huge{\textbf{On Location Relevance and Diversity in Human Mobility Data}}\footnote{The final version of this work will appear in ACM Transactions on Spatial Algorithms and Systems 7, 2, Article 7 (October 25, 2020), 38 pages, with DOI: 10.1145/3423404. Please refer to the final version for future citations.}   
\end{center}

\vspace*{1cm}

\begin{tabular}{l}
\large MARIA LUISA DAMIANI, \small{Università degli Studi di Milano, \url{maria.damiani@unimi.it}} \tabularnewline
\large FATIMA HACHEM, \small{Università degli Studi di Milano, \url{fatme.hachem@unimi.it}} \tabularnewline
\large CHRISTIAN QUADRI, \small{Università degli Studi di Milano, \url{christian.quadri@unimi.it}} \tabularnewline
\large MATTEO ROSSINI, \small{Università degli Studi di Milano, \url{matteo.rossini@unimi.it}} \tabularnewline
\large SABRINA GAITO, \small{Università degli Studi di Milano, \url{sabrina.gaito@unimi.it}} \tabularnewline

\end{tabular}

\vspace*{0.7cm}

\begin{center}
{\large \textbf{Abstract} } 

\end{center}

\vspace*{0.5cm}
The theme of human mobility is transversal to multiple fields of study and applications, from ad-hoc networks to smart cities, from transportation planning to recommendation systems on social networks.
Despite the considerable efforts made by a few scientific communities and the relevant results obtained so far, there are still many issues only partially solved, that ask for general and quantitative methodologies to be addressed. A prominent aspect of scientific and practical relevance is how to characterize the mobility behavior of individuals. In this article, we look at the problem from a location-centric perspective: we investigate methods to extract, classify and quantify the symbolic locations specified in telco trajectories, and use such measures to feature user mobility. A major contribution is a novel \emph{trajectory summarization} technique for the extraction of the locations of interest, i.e. \emph{attractive},  from symbolic trajectories. The method is built on a density-based trajectory segmentation technique  tailored to telco data, which is proven to be robust against noise. To inspect the nature of those locations, we combine the two dimensions of location attractiveness and  frequency  
into a novel  location taxonomy,  which allows for a more accurate classification of the visited places. 
Another major contribution  is the selection of suitable entropy-based metrics for the characterization of single trajectories,  based on the diversity of the locations of interest. All these components are integrated in a framework utilized  for the analysis of 100,000+ telco trajectories. The experiments show how the framework manages to dramatically reduce data complexity, provide high-quality information on the mobility behavior of people and finally succeed in grasping the nature of the locations visited by  individuals.
\vspace*{0.7cm}


\section{Introduction}
Human mobility data are key to understanding the relation of human beings with their natural, social, and built environments, for example, the use of space and time in their everyday life, which locations they visit, with what frequency and for how long.  Recent research has shown that individuals tend to visit recurrently few locations and only sometimes divert from their habitual trips to visit new locations \cite{barabasi2008}. An interesting question, of both scientific and practical relevance,  is how  to quantitatively characterize the individual mobility behavior.  

Human mobility data commonly take the form of trajectories reporting the individual location history. From a data modeling perspective, an important class of trajectories for the study of human mobility are the \emph{symbolic trajectories}. Unlike \emph{spatial} trajectories, where  
location data is collected at very fine temporal scale over a continuous space and thus the missing points in between  consecutive samples can be estimated by interpolation, symbolic
 trajectories 
are defined over a discrete space consisting of a finite set of points  $P=\{p_1,..,p_k\}$ in an embedding space.  
Given  a dictionary $L$ and a bijection $m: L \rightarrow  P$, a symbolic trajectory is a sequence of timestamped (symbolic) \emph{locations}: \begin{displaymath}T=(l_1,t_1),..,(l_n,t_n), \hbox{ with }  l_i \in L \end{displaymath} 
Locations are  spatially sparse and temporally irregular, therefore the sequence cannot be modeled as continuous trajectory or symbolic time series. 
Despite the simplicity of the model, symbolic trajectories can be effective in a variety of scenarios, to represent, for example, the movement indoors, e.g., \cite{OGC}, series of POIs, e.g., \cite{2017vldb}, series of ill-defined regions in spatial trajectories, e.g., \cite{Damiani:2014}.
A class of trajectories of major importance for the study of human mobility, which can be readily modeled as symbolic trajectories,  are those built on the call detail records (CDR) of mobile phones. 
CDRs report the communication activities of mobile communication subscribers as series of geo-referenced \emph{events}, i.e., voice call start/end, text message, data upload/download, collected by mobile operators for billing purposes. 
The importance of CDR data (\emph{game-changing data in the last decade} according to \cite{2018barbosa}), is substantially due to the significance of the user base, i.e., very large sets of individuals monitored in their daily life.   
In this work, we rely on this data we refer to as \emph{telco trajectories}, for the characterization of human mobility.%

The theme of human mobility modeling and understanding  spans multiple fields of study and applications. 
Two prominent research streams, closely related to our work, are  \emph{urban computing} and \emph{human mobility modeling}. 
Urban computing  focuses on the development of computational methods supporting the integration and analysis of heterogeneous data generated in urban spaces. The ultimate goal is to tackle major issues of interest for the city (e.g. traffic congestion, air pollution, tourism)
\cite{Zheng2014,2015csur}.  To exemplify, Zheng et al. in \cite{Zheng2011} presents a pioneering system for the recommendation of friends and locations in a city, built on methods for the detection and aggregation of locations of interest present  in  spatial trajectories.   
In contrast,  human mobility modeling focuses on the definition of  metrics and mathematical models capable of explaining and predicting the mobility behavior, e.g., \cite{barabasi2008,2018barbosa,2019kdd,2011levi,2012sim}.  For example, a fundamental result, drawn principally from the analysis of CDR data \cite{barabasi2008}, regards the probability distribution of the number of times locations are visited. 
In particular, such distribution is heavy-tail, with few locations accounting for more than 50\%  of visits, a limited set
of locations visited occasionally, while the long tail accounts for a large number of locations visited rarely or even once.  
While this is a well established characteristic of human mobility, on which there is a broad consensus in literature, a question that is still debated is how to characterize the mobility behavior of individuals. Indeed,  human mobility  has multiple facets. For example, in the related literature, one of the discriminating features is the distance covered by individuals. 
In this work, we investigate the problem from a different and location-centric perspective.

\noindent

\subsection{A location-centric view } 
An important feature of human mobility is related to the number of locations a person visits, namely, stationary people tend to frequent few locations, while people with high propensity to mobility  likely  visit  a higher number of  locations.  However, locations are not equally significant, e.g., some locations are accidental, frequented by chance, or represent noise, and thus negligible.
Therefore, simply counting the number of  locations 
may result in a coarse measure. A  different direction is to  quantify the locations that appear of major relevance for the user. 
Key challenges posed by this view are discussed in the following. 
\paragraph{Location relevance.}
In the literature on human mobility modeling, the relevance of locations  is 
 commonly related to
their frequency: the higher the number of times the location appears in the trajectory, the more relevant the location is. 
 For example, in  the seminal work \cite{barabasi2008},  the most visited location (likely home) is given rank 1, the second (likely, work place)  rank 2, and so on.  
 We refer to this class of relevance models as \emph{frequency-based}.

 A major shortcoming of frequency-based models is the semantic mismatch between frequency and importance: the locations appearing sporadically in the trajectory are classified irrelevant, regardless of the actual significance for the user, e.g. the location of an important event, while transient locations may be qualified as relevant, despite the marginal interest of the person, e.g. the bus station for a commuter. 
 
  A different viewpoint, complementary to the frequency-based perspective,  relates relevance to staying time, namely,  a location is relevant if the individual spends some significant time in it, regardless of the number of times the location is visited.  
%
%
We refer to this class of  models as \emph{attractiveness-based}. 

In this work, we hypothesize that  frequency and attractiveness are two distinct and interrelated features of location relevance. We focus, in particular, on the latter class of models.
Attractiveness-based models
are at the basis of the array of techniques developed for stop and POIs detection in spatial trajectories, e.g., \cite{Zheng2011,Parent2013,Aronov2015}. 
Unfortunately, those techniques make strong assumptions on movement, in particular on the continuity of  movement, while, in telco trajectories, the locations are sparse, with long temporal gaps between consecutive locations and with noise (see Section 3).  That raises the question of how to develop a method tailored to the discrete nature of telco data, and more in general, to symbolic trajectories. 

\noindent
\paragraph{Location diversity.} 
A natural property of locations is their diversity.
\emph{Diversity metrics} quantify the heterogeneity of a set  consisting of elements of different type (\emph{population}). While a simple measure is the count of types, more sophisticated metrics are  used in practice. The notion of diversity is key in innumerable fields, including biology, economy, demography, information theory. For example,
diversity can be used  in ecology to measure  the biodiversity of a geographical area, i.e. diversity of species; in economy, the economic variety of a region, i.e. diversity of companies with respect to their products; in demography, the racial heterogeneity in various regions. In this work, we employ  the concept of diversity to characterize the heterogeneity of relevant locations in a trajectory. We refer to that measure as \emph{location diversity}. As the number of locations as well as the number of visits vary significantly among individuals, we hypothesize that location diversity can act as discriminant for the  categorization of mobility behaviors.  

Indeed, there exists a large variety of diversity indices in the literature, e.g., Shannon-Wiener and Simpson indices, to cite a few. These indices, however, lack a unifying ground, and thus are hard to compare. Moreover, these measures do not increase linearly with the number of types, and thus are of difficult interpretation.
The challenge is to find a metric that is suitable for the problem at hand.   

\subsection{Analytical framework and contributions}
To address the above challenges,  we
develop an analytical framework 
supporting location \emph{relevance} analysis and  \emph{diversity} analysis.
%
\emph{Relevance analysis} targets the discovery and classification  of attractive locations from telco trajectories. 
Key component is a cluster-based trajectory segmentation technique, conceptually rooted in \cite{dami2018} and  tailored to symbolic trajectories. The outcome is  a set of \emph{summary trajectories}. 
The term 'summary'  
is used  to emphasize that 
the purpose of the method is to extract  key symbols from symbolic trajectories, in  analogy with text summarization methods in information retrieval. Locations are then classified based on the two criteria of attractiveness and frequency.
%
\emph{Diversity analysis} targets the quantification of location
diversity in summary trajectories.  The approach relies on the use of different entropy-based metrics, enabling  the homogeneous quantification of location diversity in terms of 'number of types'. These metrics are used to characterize the individual mobility behavior from a location perspective.
 %
\paragraph{Summary of contributions.} This article significantly extends the preliminary version  in \cite{2019Damiani} to address a few open questions, focusing in particular on the  consolidation, enrichment and validation of the analytical framework, built on the trajectory summarization technique. 
In addition, we report a richer set of experiments on a significantly larger dataset of telco trajectories. More specifically:
\begin{itemize}
	\item We  contrast the  summarization technique, referred to as \emph{SeqScan-d},  
	with a  baseline segmentation method for the 
 compression of symbolic trajectories  based on Run-Length Encoding (RLE) and data filtering. 
	While the two methods use the same set of parameters, we show that noise  sensitivity makes the difference. In particular, our technique enables a more compact representation, while
	 reducing the loss of 
	location types. 

	\item We contrast the locations in summary trajectories, i.e., the   attractive locations, with the frequent ones.  We show that frequency and attractiveness are two distinct, though interrelated location features and propose a location taxonomy based on them.  
    Specifically, we  distinguish four classes of  locations, labeled, respectively:  \emph{significant} (frequent and attractive),  \emph{transit} (frequent and not attractive),  \emph{sporadic}  (not frequent but attractive), \emph{insignificant } (not frequent and not attractive).

	\item We present an extended framework for quantifying the heterogeneity of trajectory locations. 
	In particular, we introduce: (a) the notion of \emph{location diversity profile}, grounded on recent work in biodiversity and ecology. It provides a systematic organization of the diversity concepts, enabling a multi-level description of location heterogeneity, expressed in terms of number of location types. In particular, the true diversity of order 1 and 2, built respectively on the Shannon-Wiener and Simpson diversity indices, are measures  that are differently sensitive to the number of occurrences (or type abundance) and, as such, provide different facets of the user's behavior.
	(b) An estimation of the \emph{entropy rate}, an order-dependent measure of how the entropy varies with the number of symbols (i.e. time). 
	
	\item
	We evaluate the utility  of summary trajectories. Specifically, we
 show that summary trajectories preserve major statistical properties,  in particular, the location rank distribution commonly utilized for the study of human mobility, e.g. in \cite{barabasi2008}.   
	In addition, we test the analytical framework on a dataset of anonymized 100,000+ telco trajectories (superset of the dataset used in \cite{2019Damiani}). Although larger datasets of telco trajectories are reported in literature,  e.g. \cite{2019Samet}, we consider that size in line with the state-of-the-art on research in human mobility modeling \cite{barabasi2008}.

\end{itemize}

To our knowledge, this is the first approach to the analysis of the mobility behavior built on the concepts of location relevance and diversity.  
If taken alone, the effectiveness  of location diversity metrics can be easily compromised if data are not properly filtered and organized. In this sense, relevance and diversity analysis are tightly interrelated.

The rest of the paper is organized as follows. Section 2 provides further details on related work, while Section 3 introduces the dataset used in the work along with the architecture of the analytical framework.  Section 4 focuses on relevance analysis and Section 5 on location diversity metrics. Experiments are reported in Section 6 to evaluate the summarization technique and to analyze location relevance and diversity at population scale along with  summary trajectories utility. 
 Section 7 reports a brief discussion of results. The conclusive Section 8 completes the paper.

\section{Related work}
\label{rw}

In this section, we overview major research streams related to the discovery and analysis of locations in  mobility data. We start from the methods for the discovery of core locations, rooted in urban computing, then we move to the analysis of frequent locations, focusing in particular on the statistical properties  and the metrics used to characterize the individual mobility. 

\subsection{Location  discovery techniques}
The goal of this class of techniques is to extract from a trajectory the sequence of locations that are visited by the individual (stops/places/stay points). In literature, the trajectories of concern are commonly of spatial type, thus consisting of coordinated points, while   
the  problem is formulated in terms of  trajectory segmentation, 
namely  to find the sub-trajectories (\emph{segments}) satisfying certain conditions. Two major classes of solutions are the  attribute-centric and pattern-centric segmentation techniques \cite{Damiani2017}. Attribute-centric are called the methods partitioning a spatial trajectory into a minimum
number of segments in such a way that the movement
inside each segment is nearly uniform with respect to
some condition on movement attributes. For example, a stop can be defined as a segment along which the  speed does not exceed a threshold value and whose temporal extent is lower bounded. A variety of methods for different classes of attributes can be found in literature, e.g., in \cite{Buchin2011,movemenState,Aronov2015}. 
The second stream of research on pattern-based segmentation typically leverages clustering methods for partitioning a trajectory in segments. Clustering methods are either based on simple heuristics, e.g. \cite{Zheng2011}, or on density, e.g. \cite{Palma2008,2020weib}, or partitional, e.g. \cite{2016resheff}. Clustering-centric techniques are also used for the construction of semantic trajectories \cite{Parent2013}.

A common problem to all of these  techniques is noise sensitivity. Inevitably, real data contains noise, therefore noise cannot be neglected. To tackle  this issue,
a common strategy is to introduce some additional
constraints, for example, on the maximum number
of noise points that can be tolerated. Such a parameter 
28

is, however, hard to set, especially whenever the
sampling intervals are irregular and the clustering is
to be applied to a large number of trajectories. As a
result, these techniques are highly time consuming
and not very effective in practice \cite{Damiani2017}.
A first attempt to deal in a more systematic way
with the problem of noise in clustering-based segmentation
is represented by SeqScan \cite{dami2018}. SeqScan is a
clustering technique for the segmentation of
sequences, fully compliant with the DBSCAN model \cite{Dbscan1996}.
SeqScan identifies two classes of outliers: the
points indicating a temporary absence from the cluster
(local noise) and the points representing a
definitive departure from the cluster towards
another cluster (transition points). The segmentation
algorithm determines the sequence of
clusters along with the classified outliers \cite{Damiani2017}. 

However,  SeqScan is defined for spatial trajectories, while, in our research, we have to deal with telco trajectories. We state that the aforementioned techniques  cannot be applied straightforwardly to the discrete case, as in \cite{2019Samet}. Hence,  we propose a method that, though conceptually grounded on SeqScan and thus robust against noise and relying on clustering, is  aware of the discrete nature of telco data.

\noindent
\subsection{Statistical properties of frequent locations}
The regularity and therefore the predictability of human mobility has been widely investigated and discussed over the past decade. Despite the diversity of data, methods and metrics used, there is a shared evidence that human mobility is extremely regular and therefore predictable. The main universal law is that people visit almost daily a few locations where they spend most of their time, but also visit many other locations with very diminished regularity \cite{blondel2015survey}.

The very first research \cite{2006nature} investigated human traveling statistics by analyzing the circulation of banknotes in the United States. Based on a huge dataset of over a million individual displacements, they found that the distribution of the traveling distances decays as a truncated power law, indicating that trajectories of  bank notes are similar to L\'evy flights.  Moreover, each individual tends to return to a few frequented locations with high probability.

The seminal paper on this topic is \cite{barabasi2008}. On the basis of a dataset of 100,000 trajectories of mobile phone users over 6 months, it shows that human trajectories exhibit a high degree of temporal and spatial regularity. They found that the distribution of the distance between two consecutive calls, i.e. the displacement, is well approximated by a truncated power-law, in line with \cite{2006nature}. 
A per user analysis is also conducted by calculating the radius of gyration for each user, i.e. the deviation (mean square-error) of each of the user’s positions from the average location, interpreted as the characteristic distance travelled by a user. The distribution of the radius of gyration is a truncated power-law, too. This suggests that people frequent many different locations, but the trajectories  are bounded by the characteristic user's radius of gyration. The mechanism underlying this behaviour is given by the propensity of people to return to their favourite locations over and over. To explore whether individuals return to the same location over and over, the probability of finding a user at a location was measured and plotted against the frequency rank. A linear relationship was shown,  independent of the number of locations visited by the user. The conclusion was that people devote most of their time to a few locations, although spending their remaining time in 5 to 50 places, visited with diminished regularity.
This work paved the way for a long track of researches and papers aimed to confirm, complement, enrich, quantify and eventually model this fundamental law of human mobility which was summarized in a few surveys such as \cite{blondel2015survey, 2015simi,2015csur,2019kdd}.

We mainly refer to the works which provide an estimate of this behavior such as \cite{2013physica} which found that the average number of frequently visited locations is 2.14 and that 95\% of the users visit frequently less than four locations. A complementary result is provided in  \cite{cuttone2018understanding} which found that individuals visit on average 200 unique locations in 53 weeks, of which 70\%  are visited only once. In this study a dataset from smartphones for more than
800 students at the Technical University of Denmark  was leveraged. The data sources are heterogeneous and
include GPS location, Bluetooth, SMS, phone contacts, WiFi, and Facebook friendships. The location data were collected by the smartphone with frequency of one sample every 15 minutes and the
location is determined by the best available provider.
Other works interpret regularity in visiting over and over same locations as predictability to find the user over there. For instance, Song {\em et al.} \cite{2010barabasi} studied the predictability of human trajectories derived from the estimated entropy of the mobile phone data. The predictability is centered around $93\%$ over a large population, independently of the size of the area covered by individuals' mobility or other demographic factors. Probably, the high predictability is obtained based on low resolution positioning data since the average size of a location is roughly 3 $km^2$. For higher resolution positioning data such as the GeoLife dataset, Lin and Hsu \cite{LinUbiComp12} showed that a high predictability is still present at fine spatial/temporal resolutions. However, they observed an invariance between the predictability and spatial resolution. In other words, we cannot obtain a high prediction accuracy and spatial precision simultaneously.

As regularity represents the main statistical property of human mobility, and has been also observed in the dataset we use here, we consider and investigate whether the summarization algorithm can preserve it in the summary trajectories.

\subsection{Mobility metrics}
We provide further details on mobility metrics. The major metrics  utilized for quantifying human mobility are primarily defined for discrete trajectories. 
Most of these metrics are distance-based.
For example, the \emph{jump length} \cite{2006nature} measures the distance between two consecutive locations,  
%
%
while a related metric is the Mean Square Displacement \cite{2018barbosa},  measuring the extent of the  exploration area, defined as:
$MSD(t)=\left\lvert\textbf{r}(t)-\textbf{r}_0\right\rvert^2$, 
where $\textbf{r}_0$ is the vector marking the initial point of the trajectory,  \textbf{r}(t) measures the position at time $t$.  

A major metric is the \emph{radius of gyration}, defined  as the characteristic distance travelled by the user in a time interval.   
Rooted in physics and engineering, this measure has been applied to human mobility in \cite{barabasi2008}, and defined as:
$ r_g= \sqrt{\frac{1}{N}\sum_{i=1}^N\left\lvert\textbf{r}_i - \textbf{r}_c\right\rvert^2}$, where $\textbf{r}_i$ are the coordinates of the $i$-th point of the trajectory of length $N$, and  $\textbf{r}_c$ the centroid of the set of $N$  points.  The radius of gyration  has been used  to discriminate between routinary users (\emph{returners}) and highly mobile users (\emph{explorers}). 
This measure presents, however, a major limitation. For example, if an individual spends most of the time at home and work,  and home and work are far from each other, the value $r_g$ will be very high, even though the behaviour is routinary, in clear contrast with the intended purpose of the metric. A slightly different metric trying to overcome this issue is the \emph{k-radius of gyration}, denoted   $r_g^{(k)}$\cite{2015simi}. It is defined as the radius of gyration computed over an individual's $k$ most frequently visited locations. 
If  $r_g^{(k)} \simeq r_g$ for a given $k$,  it means that the radius of gyration is dominated by  the  $k$ most visited locations, conversely, if $r_g^{(k)} \ll r_g$, the $k$ locations are not sufficient to characterize the mobility. In this sense, the value of $k$ can be interpreted as a  mobility indicator. 

The alternative  class of metrics, not relying on distance, are those based on entropy.
The model in \cite{2010barabasi} is built on the measure of the \emph{entropy rate}, which depends not only on the frequency of visitation, but also on the order in which the locations are visited.
Even if the entropy rate has been introduced as a feature to estimate the predictability of the trajectories, it can be a useful indicator 
to analyze the mobility behavior; we refer to entropy rate for that specific purpose.


\section{Data and system architecture: overview}
\label{data}
A natural starting point is to describe the nature of empirical data used for this
study. Afterwards, we describe the  architecture of the analytical framework and introduce some basic notation.

\subsection{Telco data} 
\begin{figure}[h] 
	\centering
	\includegraphics[width=0.6\textwidth]{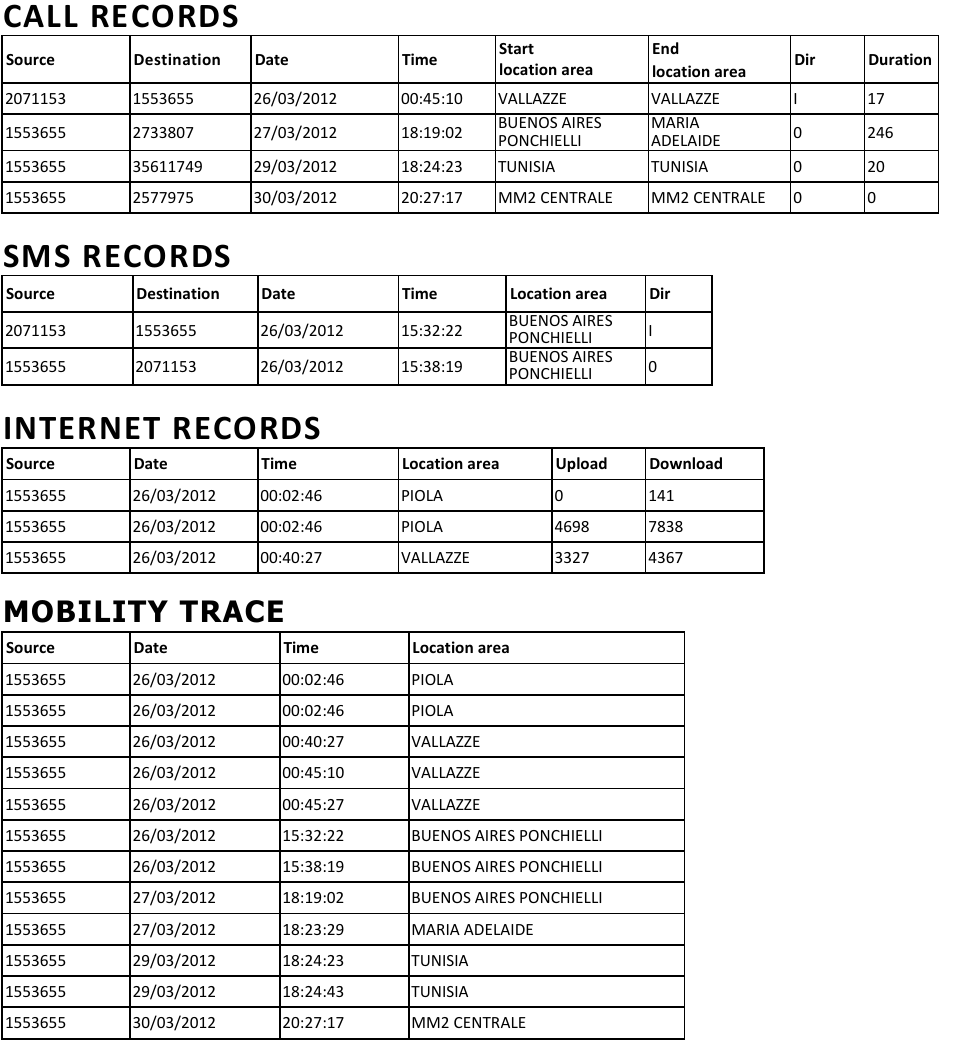}
	\caption{A fragment of CDR data}
	\label{excdr}
\end{figure}

The CDR dataset used in this work is provided by a major mobile operator in Italy. The dataset covers the city of Milan plus a few surrounding districts, over a period of 67  days, from March to May 2012.  It contains 100,000+ trajectories of anonymized users. The location information is given at the spatial granularity of \emph{Location Area}, where a  Location Area is a set of one or more base stations,  grouped together by the mobile operator, and univocally identified by a label. Figure \ref{excdr} illustrates a few records on phone calls, text messages and internet data communication, respectively.  The last sample reports the trajectory combining the records  associated to a random user. 

\begin{figure}[h] 
	\centering
	\includegraphics[width=11cm]{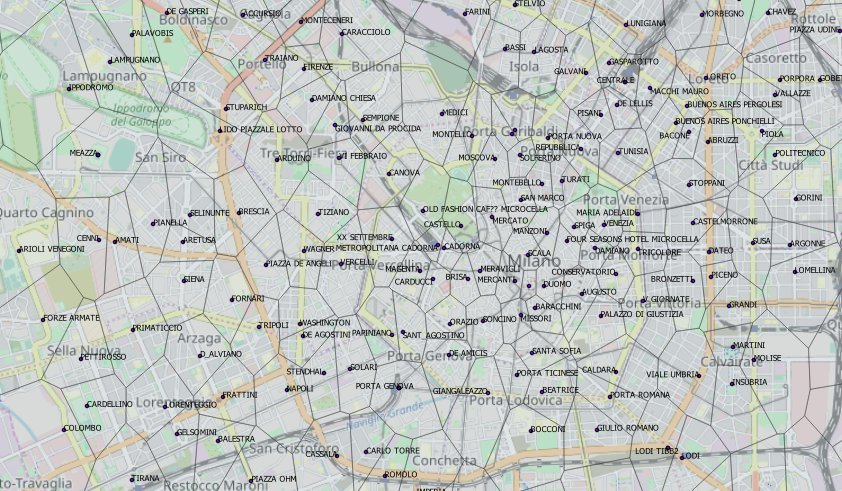}
	
	\caption{A fragment of the Voronoi diagram obtained from the estimated centroids of Location Areas in Milan}
	\label{vor}
\end{figure}

Cells and Location Areas  coordinates are not available. However, in previous work, it was estimated that  75\% of the Location Areas in Milan are smaller
than 1 square kilometer
and concentrated downtown, whilst the largest regions, over
4 square kilometers, are in the suburbs. Figure \ref{vor} shows a fragment of  the Voronoi polygons used to approximate Location Areas.  The set of representative points for the Location Areas forms the telco space.  We refer the reader to  \cite{2018sab, PAPANDREA_comcom_2016} for further details on the dataset.

Abstracting from the specific dataset,  telco trajectories have some important characteristics: 
\begin{itemize} 
	\item Sequences of identical locations.  
	Locations denote regions of space. 
	Therefore, as the user's position is matched against the closest base station, it may  happen that consecutive locations are identical.  For example, a phone call started and ended at home or in its proximity will generate two records reporting the same location. Notably, that does not happen in other kinds of trajectories, such as GPS and trajectories of check-in data, where consecutive locations are very unlikely  identical, either for technological reasons (signal characteristics) or for the nature of movement (e.g. a check-in is typically performed once).  
	\item CDRs are only generated  when phones are actively involved in a voice call, text message or internet access. Therefore large temporal gaps exist between consecutive locations. 
	Moreover, trajectories can contain bursts of  events, often related to user's activity on the internet (data upload and download), possibly interleaved by  long periods of inactivity.  
	The result is  a highly inconsistent
	temporal frequency, which may confound the mobility analysis \cite{2018barbosa}. 
	\item  The locations reported in CDRs can be noisy because of signal fluctuation in the network coverage \cite{2013physica} which can lead to ping-pong handovers between neighboring cells \cite{2003telco}. In addition, users can experience brief absences from the locations where they regularly stay, e.g. home. \cite{dami2018}. 
\end{itemize}

\subsection{The analytical framework architecture and notation}

\begin{figure}[h] 
	\centering
	\includegraphics[width=1\textwidth]{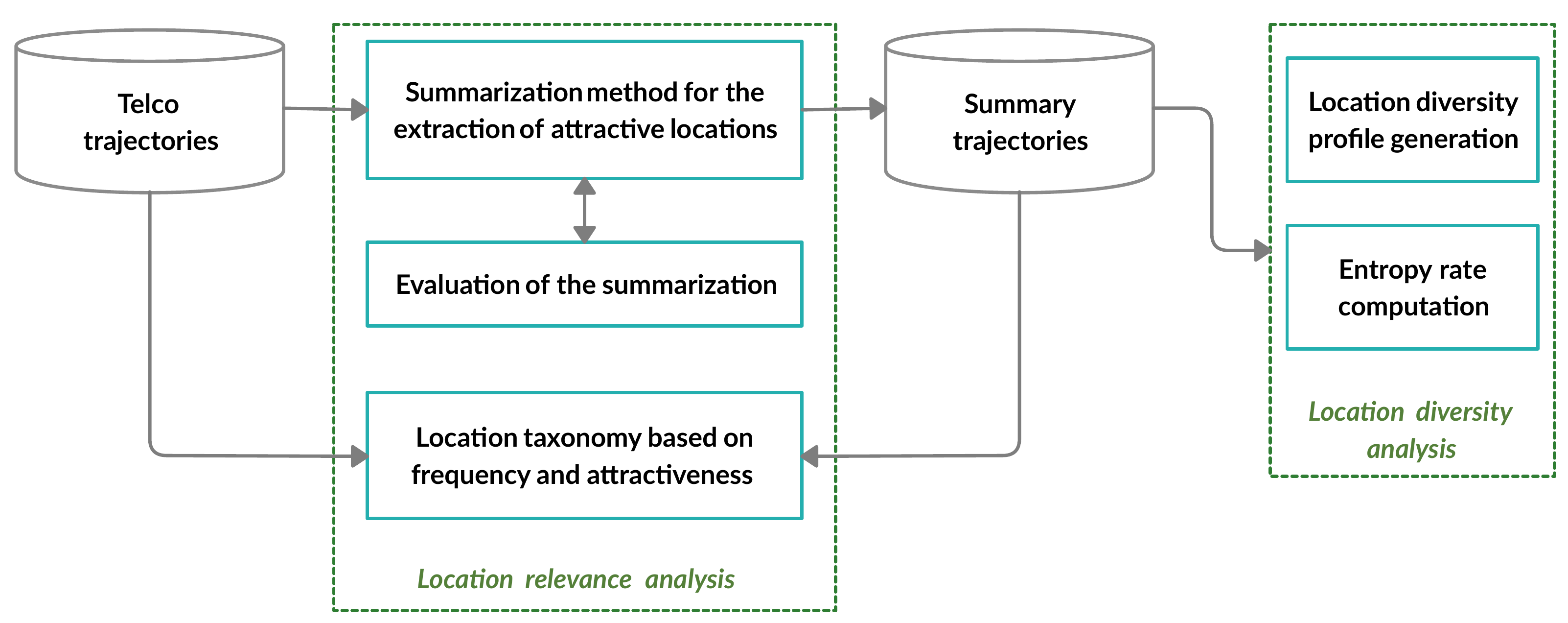}
	
	\caption{General architecture of the analytical framework: the datasets and the two building blocks for relevance and diversity analysis}
	\label{arch}
\end{figure}

Figure \ref{arch} displays the functional architecture of the framework. It consists of two building blocks supporting relevance and diversity analysis, respectively. In particular, given a dataset of telco trajectories,  relevance analysis  accomplishes two main tasks: (a) identification of key locations through trajectories summarization. The quality of clustering is assessed using internal indicators. (b) Classification of relevant locations based on frequency and attractiveness.  
The second building block is to characterize every single trajectory as a whole using the set of entropy-based indicators, specifically the location diversity profile and entropy rate estimation. The metrics are used for the classification of trajectories at population scale.


	

\subsection{Notation}
 Before proceeding, we introduce some basic notation.
  The dictionary $L$ indicates the set of locations, more specifically  the names  of the Location Areas in our dataset. Note that, throughout the paper, the terms ``location'' and  ``symbol'' are used  interchangeably.  The discrete space of coordinated points $P$ in the Euclidean plane contains the centroids $p_1,.., p_n$ of Location Areas. Thus every location has both a symbol and coordinates. The set of locations constitute the \emph{telco space}.
 
A telco trajectory $T$ is a sequence of timestamped locations in $L$, specifying where the communication events took place. If we omit the event because irrelevant in this work,  we can  simply represent a telco trajectory $T$ as a symbolic trajectory: $T=(l_1,t_1),..,(l_m,t_m)$ of length $|T|=m$ with $l_i \in L$. The symbols appearing in the dictionary represent (location) \emph{types}, while the symbols appearing in a trajectory, (location) occurrences.  We denote with $D=\{T_1,..,T_z\}$, the \emph{telco dataset}.   

A telco trajectory typically contains multiple occurrences for the same location. For example, the weekly trajectory of an individual typically contains several occurrences of 'home'. The \emph{frequency of visit} is the number of times a location appears in a trajectory.
Finally, we assume that every trajectory is univocally associated with a single user. 
The basic notation is summarized in Table  \ref{table:notation}. 
\begin{table}[h]
	\caption{Notation} 
	\centering 
	\small{
		\begin{tabular}{c  c } 
			\toprule
			$ T, D$& Trajectory, set of trajectories\\
			
			$\widehat{T}, \widehat{D} $ & Summary trajectory, set of summary trajectories \\
			L, l &  Location dictionary,  location \\
			
			$w(l,j,T), W(l,T)$&  Weight of an occurrence/symbol \\
			$\delta$ & Clustering parameter: weight threshold\\
			$N$ & Clustering parameter: occurrences threshold \\
			$H$ & Shannon-Wiener diversity index\\
            $S$ & Simpson diversity index\\
			$R$ & Richness diversity index\\
			$TD_H$, $TD_S$  & Location true diversity of order 1 and 2\\
			$S_{rate}$ & Summarization rate\\
			$Q$ & Summarization goodness\\
			$H_r$ & Entropy rate\\
			\bottomrule
		\end{tabular}
	}
	\label{table:notation} 
\end{table}

\section{Location relevance analysis: models and methods} 
In this section, we describe the technique for trajectory summarization.  We begin introducing a baseline technique for the lossy compression of telco trajectories based on RLE. Next, we detail our segmentation method, \emph{SeqScan-d}, and present a few indicators for the evaluation of segmentation. Finally, we introduce the taxonomy proposed for location classification.

\subsection{Trajectory summarization: baseline}

Broadly, the operation of trajectory summarization  extracts from  a symbolic trajectory a series of temporally annotated locations, qualified as \emph{attractive}. A first question concerns the granularity of those locations. For example, one could hypothesize that individuals commonly stay in large regions, e.g. in our case groups of nearby Location Areas, before moving to some other region.  In that case, the attractive locations could be detected by applying some form of spatio-temporal segmentation (e.g. \cite{dami2018}),  thus ignoring the symbolic dimension of locations. Actually, there is no experimental evidence that this kind of pattern exists in the reference dataset, while it is apparent that telco trajectories typically contain sub-sequences of identical symbols. This suggests an approach to segmentation based on temporal and symbolic criteria, in which attractive locations are defined at the granularity of Location Areas. The  approach is presented in the following. 

\noindent
\paragraph{Baseline technique. }
We start introducing a baseline technique for the extraction of attractive locations,  
relying on RLE compression. 
In general, given a sequence of values, RLE simply encodes subsequences $S$ of identical values (\emph{runs}) as pairs  $(l,n)$, where $l$ is the repeating value  and  $n \geq 1$ the count. 
For example, the string $aaaccvwww$ is encoded as $(a, 3)(c, 2)(v, 1) (w, 3)$. 

\begin{figure}[h] 
	\centering
	\includegraphics[width=10cm]{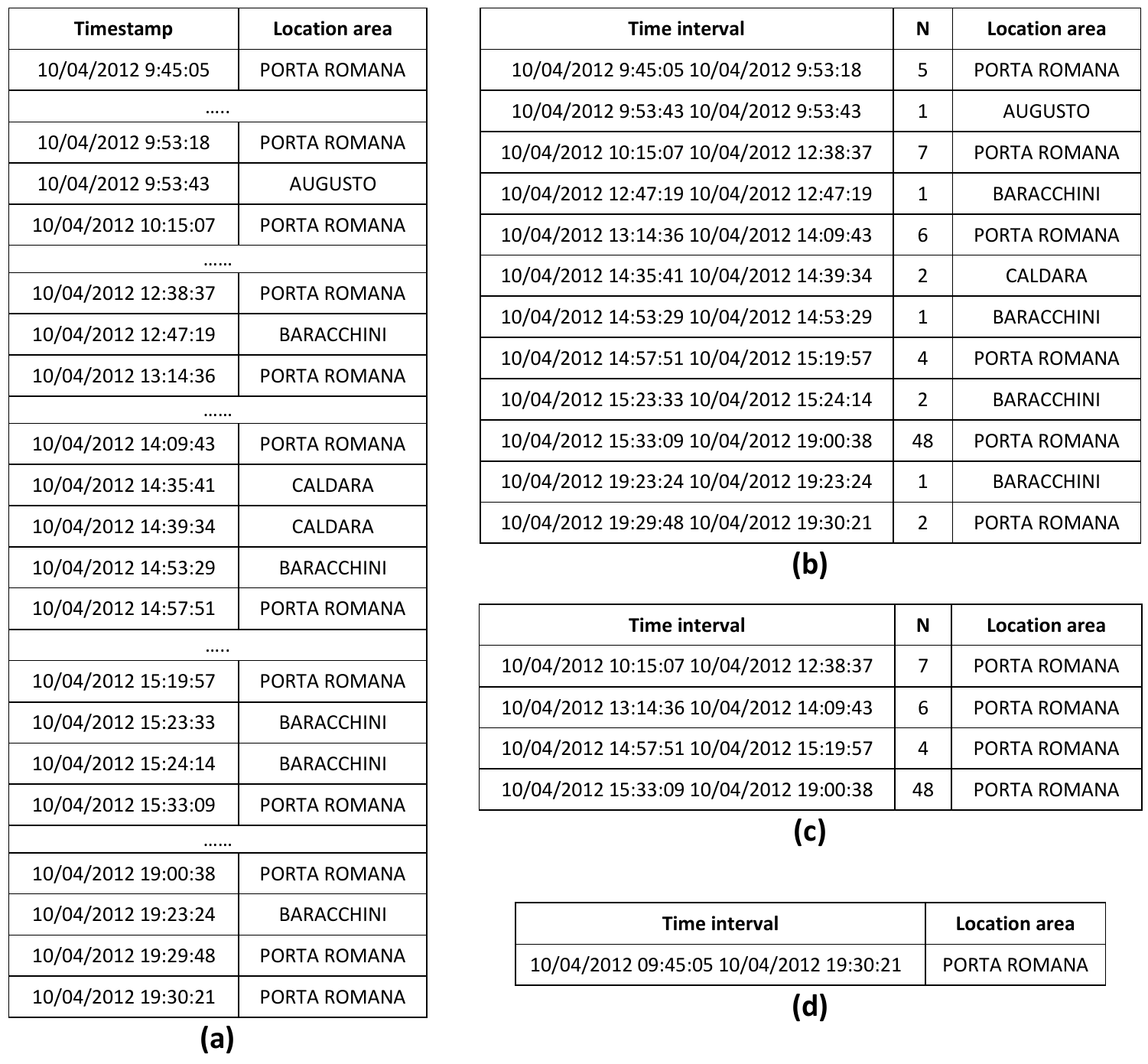}
	\caption{ (a) Fragment of the example trajectory (id=1470402); (b) portion of the  RLE trajectory; (c) portion of the RLE+ trajectory; (d) the location found by our technique}
	\label{examplePortaRomana}
\end{figure}

A straightforward application of RLE to  the compression of telco trajectories is as follows. Consider a trajectory $T$. Given a subsequence  $S$, with repeating symbol $l$, namely $S= (l, t_i)\ldots(l,t_{i+n-1})$, we encode $S$ as: \begin{displaymath}S=([t_i, t_{i+n-1}], l, n)\end{displaymath} The triple specifying the maximal time interval, the symbol and the count, i.e. $(I,l,n)$, denotes a segment of $T$. 
It can be seen that the replacement of a series of timestamps with a time period determines a loss of temporal information, while, by contrast, the symbolic information is naturally preserved.   We refer to this compressed form as \emph{RLE-trajectory}. 
\noindent
As an example,  Figure \ref{examplePortaRomana}(a)-(b) reports a fragment of a telco trajectory and the corresponding RLE-trajectory.


We are, however, only interested in a subset of locations, those that are of major importance, therefore we slightly modify RLE-trajectories, by applying the following heuristics: we  remove the segments $(I,l,n)$ where  $n < N$ or $size(I) < \delta$, with $N$ and $\delta$ indicating the minimum number of communication events (i.e. symbols), and  the minimum temporal extent respectively, for the segment to be meaningful. The former condition is to take into account the local frequency of communication events, the latter, the temporal irregularity of those events. We refer to the resulting trajectories as \emph{RLE+ trajectories} (+ to indicate the addition of constraints). 
%
%
%
Figure \ref{examplePortaRomana}(c) reports the portion of RLE+ trajectory obtained by compressing the native trajectory with $ N=4$, $\delta=16'$. It can be seen that the RLE+ representation consists of four segments. 

\begin{figure}[t] 
		\centering
        \begin{subfigure}{.485\linewidth}
            \centering
            \includegraphics[width=5.5cm]{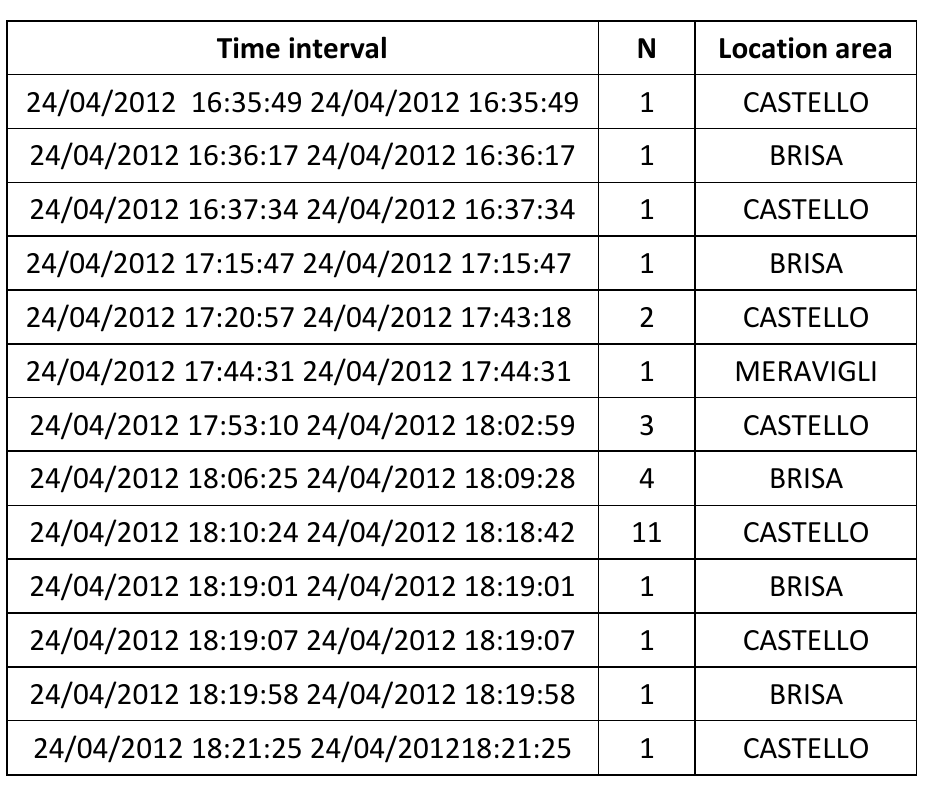}
            \caption{}
        \end{subfigure}
		\begin{subfigure}{.485\linewidth}
		    \centering
            \includegraphics[width=5.5cm]{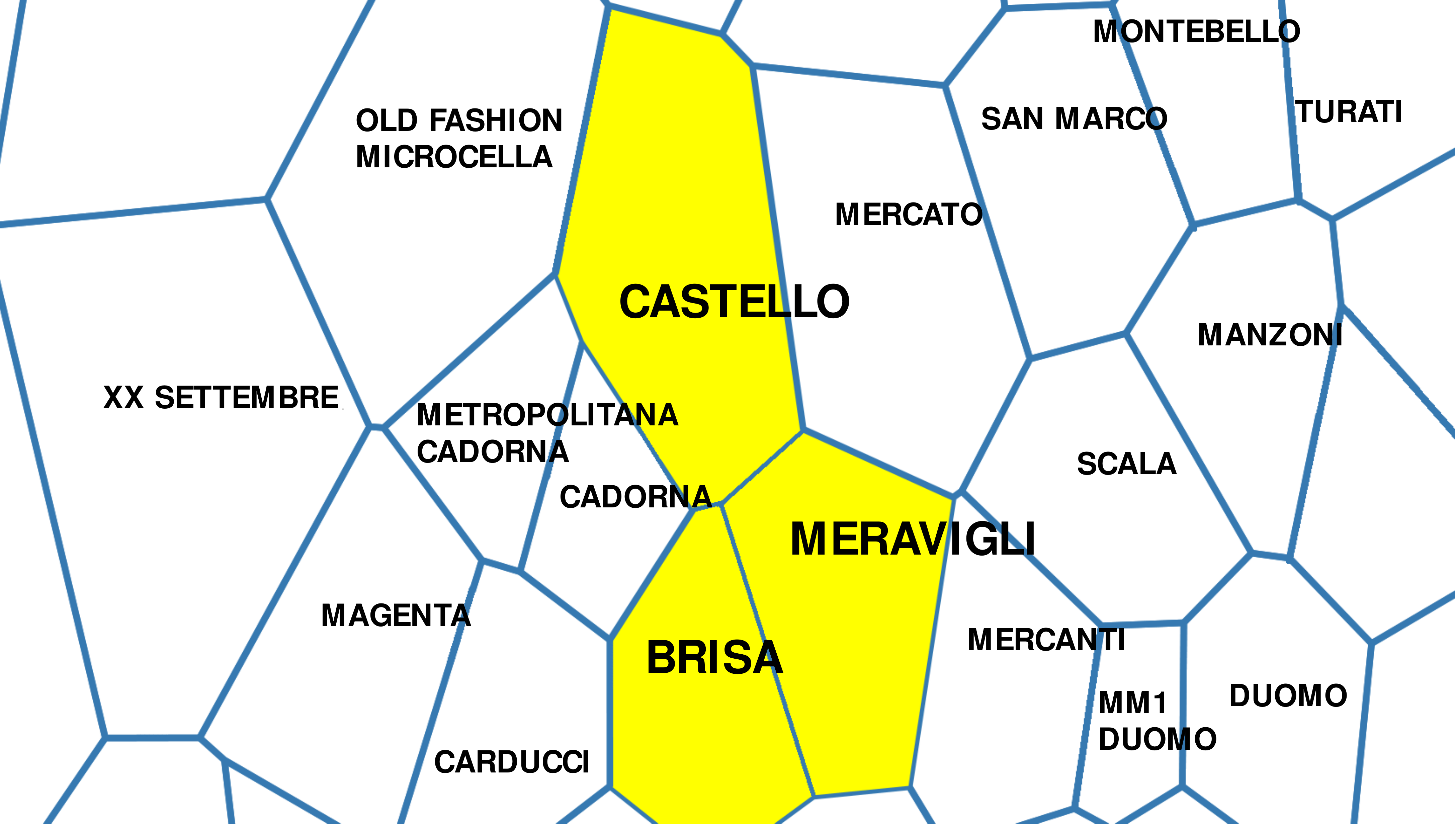}
            \caption{}
        \end{subfigure}
	\caption{(a) Fragment of RLE-trajectory (Id=1470402)  reporting the three adjacent locations displayed in (b). The portion of trajectory is displayed in the spatio-temporal coordinate system. The   $N=4, \delta=16'$, the RLE+ compression yields an empty sequence. Using our technique, the sequence is summarized in the location 'Castello' (in the period from 16:35 to 18:21 of the day 24/04/2012). The location information is thus not lost}
	\label{top20_castello}
\end{figure}

\noindent
\paragraph{Handling noise.} Unfortunately, this form of lossy compression  is sensitive to \emph{noise}. We call 'noise' the unfitting symbols in a segment.  Noise can result from signal fluctuations, or simply be the consequence of the user's movement, e.g. the temporary crossing of neighbor Location Areas. The presence of noise raises two major issues: (a) it leads to a fragmented representation of summary trajectories  confusing the analysis;
%
(b) fragmentation can determine an undesirable loss of location information.
We illustrate the latter case with the following example.  
Consider the RLE+  trajectory (fragment) in Figure \ref{top20_castello}(a). It can be seen that the user's movement in the time period falls in three spatially adjacent Location Areas (within a touristic area) (Figure \ref{top20_castello}(b)). 
Yet, the segments do not satisfy the constraints (with $ N=4$, $\delta=16'$), and thus are removed from the  RLE+ trajectory. As a result, the whole visit to this touristic area is removed. 

To reduce the impact of noise, we present a  technique conceptually built on  density-based trajectory segmentation. It requires the same parameters as RLE+, yet it
allows for a more compact representation. 
For example,  the trajectory in Figure \ref{examplePortaRomana}(a) is summarized in one segment (see Figure \ref{examplePortaRomana}(d)) covering the whole time of the visit. This result can be read in this way: the location  ('Porta Romana') is the most attractive location visited by the user  in the period from 9:45 to 19:30 of the day 10/04/2012.  
Regarding the example trajectory in Figure \ref{top20_castello}(a), it can be shown that, unlike RLE+, our technique  can find an attractive location ('Castello', in the period from 16:35 to 18:21 of the day 24/04/2012), thus reducing the location information loss.

\subsection{Density-based segmentation of  symbolic trajectories} 
\begin{figure}[h] 
	\centering
	\includegraphics[width=8cm]{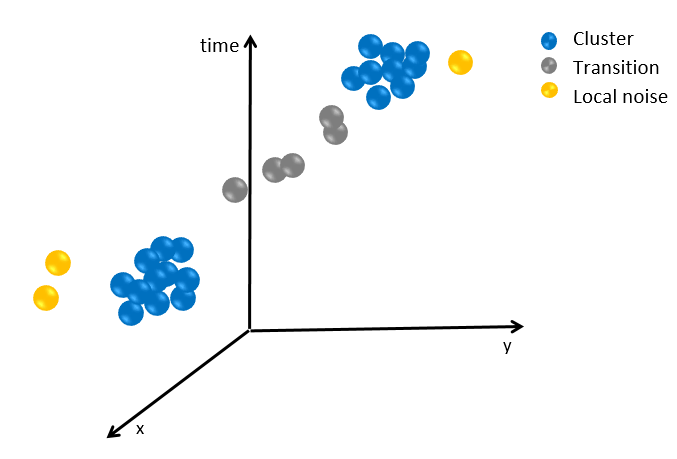}
	\caption{Density-based segmentation of a spatial trajectory \cite{dami2018}. The spatio-temporal points are classified as: cluster point, noise, transition point}
	\label{initt0}
\end{figure}

We propose an approach  that takes inspiration from SeqScan \cite{dami2018}. SeqScan partitions a spatial trajectory  in a series of temporally ordered clusters of arbitrary shape interleaved by sequences of unstructured points called \emph{transitions}. The points that do not belong to any  cluster or transition are classified as \emph{local noise}. Figure \ref{initt0} illustrates the concepts of cluster, transition and local noise.
\begin{figure}[h] 
	\centering
	\begin{subfigure}{0.485\textwidth}
	\centering
	    \includegraphics[width=.95\linewidth]{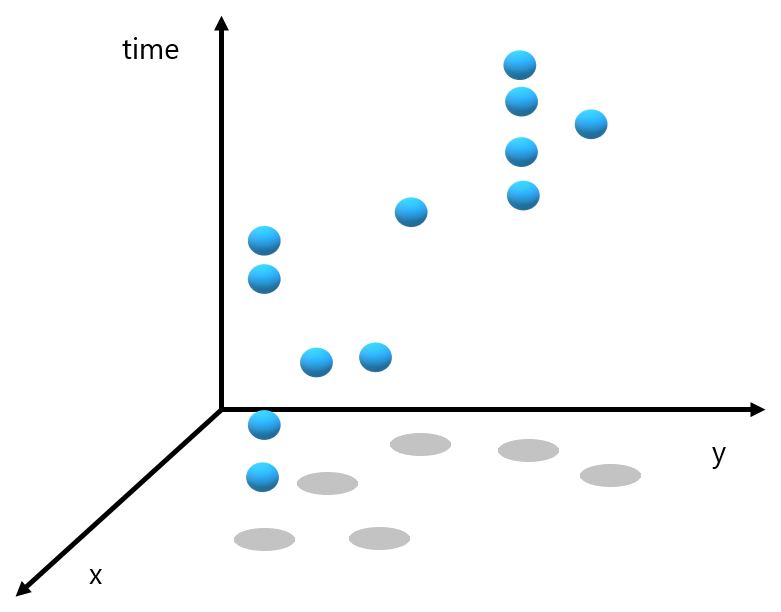}
	\end{subfigure}
	\begin{subfigure}{0.485\textwidth}
	\centering
	    \includegraphics[width=.95\linewidth]{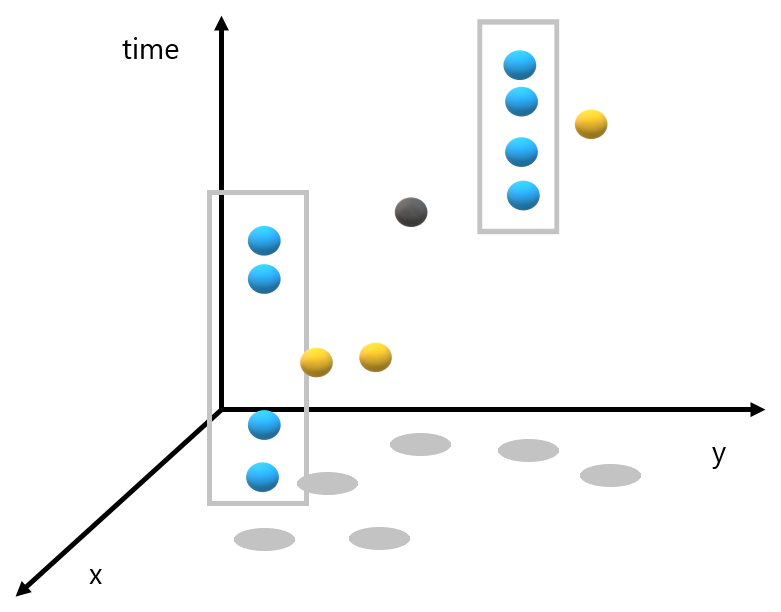}
	\end{subfigure}
	\caption{(a) Telco trajectory with 6 different locations (on space) and 12 occurrences (in space-time); (b) The rectangles contains two groups of occurrences for the same symbol, representing clusters along the temporal line. There are also 3 noisy points and one transition point}
	\label{new}
\end{figure}  

SeqScan is, however, tailored to spatial trajectories,  while proven ineffective  when applied to telco  trajectories, therefore
%
we have investigated a different strategy. 
To convey the intuition, 
consider the simple sequence in Figure \ref{new}.(a). This example shows a  space of 6 locations (as points in the plane)  and  the trajectory of a user. 
It can be seen that the trajectory contains sequences of identical locations, meaning that the user is located in the same region at different times while making a phone call or accessing the Internet.  As the user moves elsewhere, two cases can occur: the user returns back to the previous location;  or the user starts frequenting some other location. This suggests a cluster-based segmentation performed over the temporal line with clusters only grouping occurrences of a unique location. Figure \ref{new}.(b) shows two clusters, few points of local noise and one point classified as transition.  A cluster is described by a segment, i.e.  a symbol and  a temporal extent. 
Our hypothesis is that a cluster identifies  the occurrence of an attractive location. We refer to this  technique as \emph{SeqScan-d} (discrete). Similar to RLE+, SeqScan-d requires the specification of two parameters $N$ and $\delta$.  
The model and the algorithm are detailed in the following.

\subsubsection{The model}
The model is built on the concept of dominant symbol. Intuitively, a \emph{dominant} symbol (or location) is a symbol that is  representative of a time period. The number of occurrences  is a natural relevance metric. 
For example, in the sequence: $(a, t_1) (a, t_2) (b, t_3) (a, t_4)$, the symbol $a$ appears 3 times in the period $I=[t_1, t_4]$, 
thus $a$ is dominant in $I$, irrespective of the presence of $b$.
However, sequences can be quite complex and long,  and different symbols can compete for the  role. 
For a more selective competition, symbols are given a weight. Such weight awards temporally correlated occurrences of the same symbol.  Initially every symbol has weight 0; afterwards, if two consecutive symbols at time $t_i$ and $t_{i+1}$ are identical, the weight of that symbol increases of a certain amount. In particular, the amount is given by the temporal distance $|t_{i+1}-t_i|$. A dominant location is  a location that is sufficiently frequented and has high weight. 
More formally, consider a trajectory $T=(l_1, t_1)..(l_n, t_n)$, defined in the interval $[t_1, t_n]$. 

\begin{Def}[\textbf{Weight function}]
We introduce:
\begin{itemize}
\item The  function computing the  weight of the occurrence of $l\in L$ at position $j$ in $T$, is defined as:
	\[
	w(l,j, T)=
	\begin{cases}
	|t_{j}-t_{j-1}|, & \text{ if } j>0 \hbox{ and }l_j=l_{j-1}=l  \\
	0,             & \text{otherwise}.
	\end{cases}
	\] 
	
	\noindent
\item 	The weight  $W(l, T)$ of a symbol $l$  over  the trajectory $T$ is given by the sum of the weights  of the  symbol occurrences:
	
\begin{equation}
	W(l, T)= \sum_{j=1}^n w(l,j, T).
\end{equation}
\end{itemize}
\end{Def}
\rm
\begin{example} Consider the following trajectory  from $t_1$ to $t_{10}$ consisting of 10 occurrences, homogeneously spaced in time of 2 time units.   
\begin{equation}
\label{eq}
T=(a, t_1) (a, t_2) (c, t_3) (a, t_4) (c, t_5) (b, t_6) (b,t_7) (a, t_8) (b, t_9) (b, t_{10})
\end{equation}
The symbols in $T$ have the following weight:  
\begin{itemize}
\item []$W(a,T)=(t_2 - t_1)=2$
\item []$W(b, T)=(t_7 - t_6) + (t_{10} - t_9)=4$
\item []$W(c,T)=0$ 
\end{itemize}
It can be seen that the symbol $b$ has the highest weight. $\diamond$ 
\end{example}


We now introduce the notion of dominant symbol. The definition is built on the auxiliary  definition of well-formed sequence.  Consider a sequence $T=(l_1, t_1)..(l_n, t_n)$,
 $N \geq 2$ and  $\delta \geq 0$.
\begin{Def}[\textbf{Well-formed sequence}]
    We say that $T$ is a well-formed sequence if the following conditions are satisfied:
    \begin{itemize}
    	\item[i.] $l_1=l_n$;  
    	\item[ii.] $W(l,T) \geq \delta $;
    	\item[iii.] $|\{ l_j \in T \mid l_j =l_1 \}| \geq N$.
    \end{itemize}
\end{Def}
Condition (i) states that  the sequence $S$ is bounded by $l$; (ii) and (iii) define lower bounds for the relevance measures.
Now, the dominant symbol can be defined as follows:
\begin{Def}[\textbf{Dominant symbol}] 
	Given a symbol $l$, we say that  $l$ is dominant in $T$ if the following conditions are satisfied:
	\begin{itemize}
	    \item[i.] $T$ is a well-formed sequence having $l$ as first symbol, i.e. $l=l_1$;
		\item[ii.] Let $(l_1, t_1)..(l_k, t_k)$ the minimal-length well-formed sub-sequence of $T$ that starts at index $1$; then $(l_{k+1},t_{k+1})..(l_n, t_n)$, if not empty, does not contain well-formed sub-sequences having $l'\neq l$ as first symbol. 
	\end{itemize}
	If $l$ is dominant in $T$, the period $I=[t_1, t_n]$ is called the \emph{dominance period} of $l$.
\end{Def}

The minimal-length well-formed sub-sequence at point (ii) represents the initial cluster of symbol occurrences. Such a cluster is then extended with occurrences of the same symbol and possibly noise, until a new cluster is found.
As a consequence of the definition, the dominant symbol 
is unique in the time frame of $T$.
\noindent
Finally we define a summary trajectory a sequence of dominant symbols. 
\begin{Def} [\textbf{Summary trajectory}]
	A summary trajectory $\widehat{T}$ is a  trajectory of length $w$: $$\widehat{T}= (I_1,l_1)..(I_w, l_w)$$ where the unit $(I_j, l_j)$ specifies the symbol $l_j$ dominant in the period $I_j$, and $I_j$ is maximal.  The dominant symbols are the \emph{attractive} locations in the native trajectory $T$.

\end{Def}

\noindent
\begin{example} 
Consider again trajectory \eqref{eq}. If $N=3, \delta=2$, the summary trajectory is:
\begin{equation}
\label{st}
\widehat{T}= ([t_1, t_4], a) ([t_6, t_{10}], b)
\end{equation}
If $N=3, \delta=4$, the summary trajectory is: 
\begin{equation}
\widehat{T}= ([t_6, t_{10}], b) 
\end{equation}
Note that the summary trajectory in (4) is obtained by specifying a more restrictive temporal constraint  than in (3) $\diamond$.
\end{example}

The condition on the maximality of the dominance period implies that the sequence is to be as long as possible, compatibly with the constraints. Summary trajectories can be straightforwardly represented using the symbolic trajectories data model in \cite{tsas2015}.
The dominant symbol identifies a cluster of occurrences, while the symbols other than the dominant symbol in a period are local noise \cite{dami2018} or \emph{absences} in that period.  Note that summarizing trajectories determines a loss of information.

\subsubsection{The algorithm}
  Given the input parameters $N$ and $\delta$, the algorithm extracts a series of dominant symbols as temporally separated clusters. The symbols of the sequence are processed one at a time. As  a dominant symbol is found, a cluster is created and becomes the active cluster. The algorithm proceeds trying to expand the active cluster, while  monitoring at the same time the emergence of other clusters.  If the active cluster is no longer expanded, and a new symbol becomes dominant, the active cluster is closed and appended to the output clusters, while a new cluster is created. The process terminates when the scan of the sequence is complete. The outcome is a series of temporally annotated 
  clusters, each representing a dominant location.  
 
\noindent
The algorithm is detailed in Algorithm~\ref{algoSeq}.
The information relevant for the processing of symbols is kept in a hash table for the symbols  of the telco space. 
For every distinct symbol $s$ of the trajectory, the tuple $(n,w,l)_s$  reports the number of occurrences, the weight and the index of the first occurrence in the portion of trajectory being processed.  
As a symbol $s$ becomes dominant, the hash table, except for the dominant symbol entry, is reset and the phase of cluster expansion starts. Upon the reading of a symbol $s'$, two cases may occur: a) if $s'$  is an occurrence of the dominant symbol, the entry is updated while the hash table is reset again, as above.  Note that the reset operation is necessary to ensure that clusters are temporally disjointed. b) If $s'$ is not an occurrence of the dominant symbol, the corresponding entry in the hash table is updated and the input constraints are checked.  Hence, if the  symbol $s'$ becomes dominant, the current cluster is closed and the pair $(I,s)$, with $I$ denoting the time interval between the first and the last occurrence of $s$, is stored as \emph{unit} of the summary trajectory. The output of the algorithm is a list of units defined over temporally separated  time intervals. 

The time complexity of the algorithm is linear with respect to the number of points in the original trajectory. The usage of the hash table guarantees 
a constant access time. 
As for the space requirement, the algorithm  keeps in memory the original trajectory $T$, the hash table $H$ and the summary trajectory $\widehat{T}$. %
The size of the hash table equals the size of the dictionary $L$. 
Large datasets of trajectories can be summarized using modern multi-core CPUs and parallel computational frameworks. Moreover, the algorithm can be used in both batch and streaming mode. 

\begin{algorithm}
	\SetAlgoLined
	\hrule
	\vspace*{0.3cm}
	\SetKwInOut{Input}{Input}\SetKwInOut{Output}{output}
	
	\Input{$T= [(l_1,t_1),(l_2,t_2),...]$ \textit{\scriptsize{//trajectory }} \\
	$N$, $\delta$ \textit{\scriptsize{//input parameters}}}
	
	\KwResult{$\widehat{T}$\textit{\scriptsize{//summary trajectory}} }
	
	$C$=$\emptyset$ \textit{\scriptsize{//Active cluster}} \;
	$H$ \textit{\scriptsize{//Hash table of $|L|$ entries}} \;
	
	\For{$(l,t)$ in $T$}{
		
		\textbf{$H$.UpdateEntry( $l$, $t$)}\; 
		\eIf{$C$=$\emptyset$}{
			\If{getsDominant($l$)}{
				$C$ $\leftarrow$ \textbf{Cluster($l$)}\;
				\textbf{$H$.ResetNonDominantSymbols($l$)}\;
			}
			
		}{
			\eIf{$l$==\textbf{dominant($C$)}}{
				\textbf{$H$.ResetNonDominantSymbols($l$)}\;
			}{
				\If{\textbf{getsDominant($l$)}}{
					\textbf{$\widehat{T}$.Add(close($C$))}\;
					$C$ $\leftarrow$ \textbf{Cluster($l$)}\;
					\textbf{$H$.ResetNonDominantSymbols($l$)}\;
				}
			}   
			
		}
	}  
	\vspace*{0.3cm}
	\hrule
		
	\caption{Summarization algorithm }
	\label{algoSeq}
\end{algorithm}

\subsubsection{Evaluation metrics}
To evaluate the segmentation, we use two measures: \emph{summarization rate}  and \emph{summarization goodness}, respectively,  defined as follows:  
\begin{itemize}
	\item  The  summarization rate  $S_{rate}(T)$ specifies the percentage of locations types (i.e. different symbols) in the native trajectory $T$  that do not appear in the summary trajectory $\widehat{T}$, in other words, the percentage of irrelevant locations. Let $R()$ be the number of location types 
	in the input trajectory. We have:  %
	\begin{equation}
	S_{rate}(\widehat{T})=1-\frac{R(\widehat{T})}{R(T)}
	\end{equation}
	%
	$S_{rate}\approx 0 $ means limited or even  no summarization; $S_{rate} \approx 1 $, a high level of summarization. 

\begin{example}
Consider the   trajectory $T$, consisting of 10 occurrences, i.e. $|T|=10$ and 3 distinct location types, i.e. a, b, c: $$(a,t_1) (b,t_2) (a,t_3) (a, t_4) (c,t_5) (b,t_6) (c,t_7) (c, t_8) (a, t_9) (a, t_{10}) $$
We have  $R(T)=3$. Assume the following summary trajectory: $$\widehat{T}= ([t_1,t_4], a) ([t_5,t_8], c) ([t_9,t_{10}],a) $$ As $\widehat{T}$ contains  only 2 distinct symbols, the summarization rate is: $S_{rate}(T)=1/3$. 
\end{example}
\noindent
	
	Finally, we define the summarization rate $S_{rate}(\widehat{D})$ for the whole dataset as the average summarization rate of the trajectories, i.e.;  $$ S_{rate}(\widehat{D})= \frac{1}{|\widehat{D}|}\sum_{i=1}^{|\widehat{D}|} S_{rate}(\widehat{T}_i) $$
	
	\item The goodness of a summary trajectory $Q(\widehat{T})$ measures the quality of clustering. 
	For the evaluation of clustering, we use the internal index proposed for the validation of the SeqScan method \cite{Damiani2016} (\emph{stationary index}). 
	Accordingly, the quality of clustering  is measured in terms of temporal density of clusters.
	In essence, the less the local noise in the period, the stronger the dominance and thus the goodness of the cluster.  
	
	In more formal terms, let $T' \subseteq T$ be a subsequence summarized in the unit $u_i=(I_i, l_i )$, with $l_i$ the dominant location in the period $I_i=[t_b, t_e]$.  We define the goodness of the unit, denoted $q(u_i)$, as the ratio of the  symbol weight in $T'$  and the length of the dominance period, i.e.,  $q(u_i)=\frac{W(l_i, T')}{(t_e-t_b)}$. Intuitively, it can be seen as the percentage of time estimated to be spent in the dominant location.
	The goodness of a summary trajectory is  the average goodness of the units $ Q(\widehat{T})= \frac{1}{n}\sum_i^n q(u_i) $, with $n=|\widehat{T}|$.  
	Finally, the goodness of the summary dataset  $Q(\widehat{D})$ is the average goodness of summary trajectories, i.e.: $$ Q(\widehat{D})= \frac{1}{|\widehat{D}|}\sum_{i=1}^{|\widehat{D}|} Q(\widehat{T}_i) $$

\end{itemize}
\begin{example} Consider trajectory \eqref{eq} and its summarization in two units, denoted hereinafter $u_1, u_2$, i.e. $\widehat{T}= ([t_1, t_4], a) ([t_6, t_{10}], b)$. We have: 
\begin{itemize}
    \item $S_{rate}(T)= 1/3$. In fact, there is only one irrelevant symbol, i.e. $c$.
    \item $Q(\widehat{T})= 5/12$, where:
    \begin{itemize}
    \item $q(u_1)= (t_2 - t_1)/(t_4 - t_1)= 1/3$
    \item $ q(u_2)= [(t_7 - t_6) + (t_{10} - t_9)] /(t_{10} - t_6) = 1/2$
    \end{itemize}
    
\end{itemize}
\end{example}

\subsection{Insights into the nature of locations: the proposal of a location taxonomy}
\label{sec:location_taxonomy}
Trajectory summarization is applied to the extraction of attractive locations from telco trajectories. 
We now turn to discuss the nature of the relationship between 
location frequency and attractiveness. 

The frequency of visiting a location in fact accounts for how many times a person is there but is unable to give any information on the attractiveness that the location has on it. There may be locations, such as the bus stop, which are often visited by a person who has no specific interest in it, but just transits through it. On the contrary, we find examples such as a museum, a place that is rarely visited but that expresses a strong cultural attraction, a place therefore detected as attractive by SeqScan-d, even if visited only sporadically. 
The synergy between the two metrics reinforces the role that a location plays in a person's life: locations that are both attractive and frequent are clearly very significant, while non attractive and infrequent ones are non-significant. 
 Thus, by combining these two dimensions, we can gain further insights into the nature of the visited locations.
 
\begin{figure}[t]
    \centering
    \includegraphics[width=5cm]{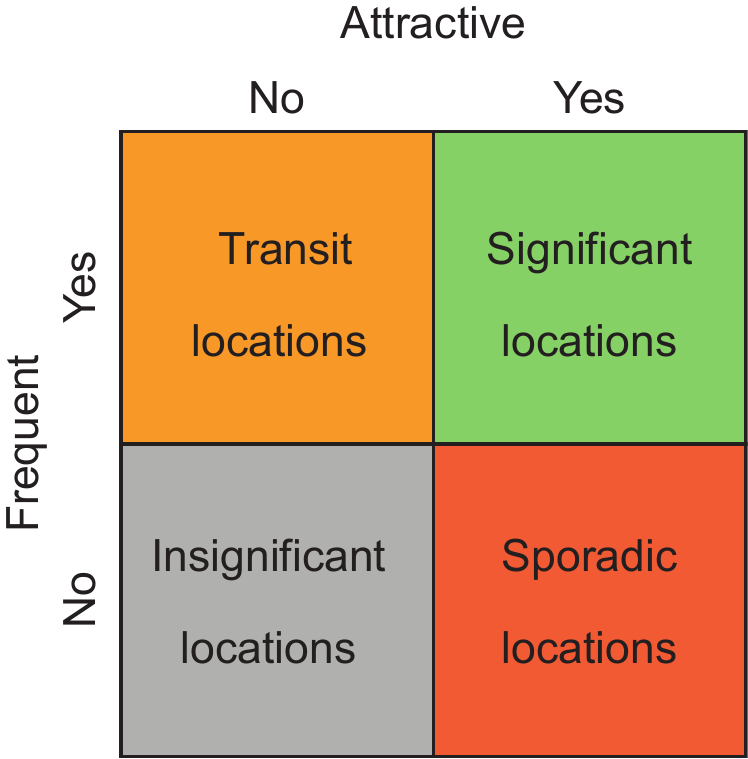}
    \caption{Location taxonomy} 
    \label{fig:label_table}
\end{figure}

 In particular, we distinguish four classes
of locations that we label:  \emph{significant (SL), transit (TL), sporadic (PL), insignificant (IL)}, respectively (in Figure  \ref{fig:label_table}). 
The semantics of these classes can be straightforwardly defined in set-based terms. For the sake of generality, assume a  generic frequency metric. Consider a trajectory $T$ and denote with $T_u$ and $\widehat{T_u}$ the two sets of distinct  locations in $T$ and $\widehat{T}$, respectively. Moreover, denote with $N_u$ the top-n frequented locations in $T$, where $n=|\widehat{T_u}|=|N_u|$, as shown in Figure \ref{fig:schema_sets}. We define: 

\begin{figure*}
    \centering
    \includegraphics{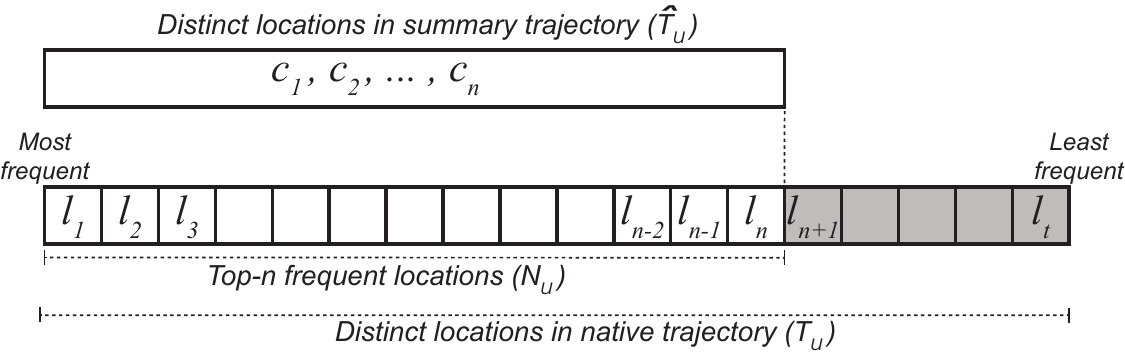}
    \caption{Schema of the sets $\widehat{T_u}$ and $N_u$ used in the relevance analysis to build the location taxonomy and to compare the frequency-based and the attractive-based approaches. Given the set $\widehat{T_u}$ of the $n$ attractive locations of the summary trajectory and the  set $T_u$ of the distinct locations in the native trajectory ordered by descending frequency, the set $N_u$ is built by considering the top $n$-frequented locations  
    }
    \label{fig:schema_sets}
\end{figure*}

\begin{itemize}
    \item  \emph{Significant} locations are those that are both frequent and attractive, i.e., $SL=N_u \cap \widehat{T_u}$;
    \item \emph{Transit} locations are those that are frequent but not attractive, for example the locations where the user passes quickly daily, i.e.,  $TL=N_u\setminus \widehat{T_u}$;
    \item  \emph{Sporadic} locations are those that are attractive but  not frequent, for example, the locations hosting some event of interest for the user, e.g., $PL=\widehat{T_u} \setminus N_u$;
    \item  \emph{Insignificant} locations are those that are neither attractive or frequent, i.e., $IL=T_u \setminus (\widehat{T_u} \cup N_u)$.
\end{itemize}

\noindent
As a result, given  an input trajectory $T$ and its summarization $\widehat{T}$, we can label each distinct location in $T$, based on the proposed taxonomy. To better illustrate the operation, we use an example.


\begin{example}  
The trajectory with ID=1470401 is summarized  with parameters $N=4, \delta=16'$. Figure \ref{fig:taxomony_example} 
shows the set  $T_u$ of locations  ordered by descending frequency, 
the set $\widehat{T_u}$ of 9 locations detected by the summarization algorithm, 
 the corresponding 
 set of the 9 most frequent locations, i.e. $N_u$. 
The location classes  $SL, TL, PL, IL$, drawn from the previous sets, are  reported along with the percentage of locations for each class. It can be seen that, in this specific case, the attractive locations amount to about 12\% of the total, nearly equally subdivided in  significant and sporadic locations. 
\end{example}
\begin{figure*}
    \centering
    \begin{subfigure}{0.485\textwidth}
    	\centering
    	{
\centering
\tiny
\begin{tabular}{|l|l|l|} 
\cline{2-3}
\multicolumn{1}{c|}{} & \multicolumn{1}{c|}{\textbf{Location area}} & \multicolumn{1}{c|}{\textbf{Frequency}}  \\ 
\hline
1                     & ABRUZZI                                     & 723                                      \\ 
\hline
2                     & LAGOSTA                                     & 129                                      \\ 
\hline
3                     & ACCURSIO                                    & 124                                      \\ 
\hline
4                     & PORTA NUOVA                                 & 79                                       \\ 
\hline
5                     & BACONE                                      & 48                                       \\ 
\hline
6                     & NIGUARDA                                    & 33                                       \\ 
\hline
7                     & PIOLA                                       & 23                                       \\ 
\hline
8                     & MM2 CENTRALE                                & 16                                       \\ 
\hline
9                     & CARACCIOLO                                  & 16                                       \\ 
\hline
10                    & STELVIO                                     & 15                                       \\ 
\hline
11                    & BELGIOIOSO                                  & 12                                       \\ 
\hline
12                    & VIA DE PREDIS                               & 12                                       \\ 
\hline
13                    & MARCHE                                      & 11                                       \\ 
\hline
14                    & STUPARICH                                   & 10                                       \\ 
\hline
15                    & MINCIO                                      & 10                                       \\ 
\hline
16                    & GORINI                                      & 9                                        \\ 
\hline
17                    & MORBEGNO                                    & 8                                        \\ 
\hline
18                    & BUENOS AIRES PONCHIELLI                     & 7                                        \\ 
\hline
19                    & STOPPANI                                    & 7                                        \\ 
\hline
20                    & PALAVOBIS                                   & 6                                        \\ 
\hline
21                    & DAMIANO CHIESA                              & 6                                        \\ 
\hline
22                    & PISANI                                      & 6                                        \\ 
\hline
23                    & GOBBA                                       & 6                                        \\ 
\hline
24                    & GASPAROTTO                                  & 5                                        \\ 
\hline
25                    & VIALE UMBRIA                                & 5                                        \\ 
\hline
26                    & LORETO                                      & 5                                        \\ 
\hline
27                    & MOLINO DORINO                               & 4                                        \\ 
\hline
\multicolumn{3}{|c|}{$\dots$}                                                                                  \\ 
\hline
38                    & MM1 DUOMO                                   & 4                                        \\ 
\hline
39                    & ROGOREDO                                    & 3                                        \\ 
\hline
\multicolumn{3}{|c|}{$\dots$}                                                                                  \\
\hline
45                    & PIAZZA XXV APRILE                           & 3                                        \\ 
\hline
46                    & CASA DEL SOLE                               & 2                                        \\ 
\hline
\multicolumn{3}{|c|}{$\dots$}                                                                                  \\ 
\hline
53                    & MAFFUCCI                                    & 2                                        \\ 
\hline
54                    & DE LELLIS                                   & 1                                        \\ 
\hline
\multicolumn{3}{|c|}{$\dots$}                                                                                  \\ 
\hline
72                    & CONSERVATORIO                               & 1                                        \\
\hline
\end{tabular}

}
    	\caption{$T_u$ {\scriptsize (ordered by frequency)} }
    	\label{fig:taxomony_example_t_u}
	\end{subfigure}
	\begin{subfigure}{0.485\textwidth}
	    \begin{subfigure}{.48\textwidth}
        	\centering
        	{
\tiny
\centering
\begin{tabular}{|l|l|} 
\cline{2-2}
\multicolumn{1}{l|}{} & \multicolumn{1}{c|}{\textbf{Location area}}        \\ 
\hline
1                     & ABRUZZI  \\ 
\hline
2                     & ACCURSIO                                           \\ 
\hline
3                     & BELGIOIOSO                                         \\ 
\hline
4                     & CHIESE                                             \\ 
\hline
5                     & GORINI                                             \\ 
\hline
6                     & LAGOSTA                                            \\ 
\hline
7                     & MINCIO                                             \\ 
\hline
8                     & NIGUARDA                                           \\ 
\hline
9                     & PORTA NUOVA                                        \\
\hline
\end{tabular}

}

        	\caption{$\widehat{T_u}$}
        	\label{fig:taxomony_example_c_u}
	    \end{subfigure}
	    \begin{subfigure}{.48\textwidth}
        	\centering
        	{
\tiny
\centering
\begin{tabular}{|l|l|} 
\cline{2-2}
\multicolumn{1}{l|}{} & \multicolumn{1}{c|}{\textbf{Location area}}        \\ 
\hline
1                     & ABRUZZI  \\ 
\hline
2                     & LAGOSTA                                            \\ 
\hline
3                     & ACCURSIO                                           \\ 
\hline
4                     & PORTA NUOVA                                        \\ 
\hline
5                     & BACONE                                             \\ 
\hline
6                     & NIGUARDA                                           \\ 
\hline
7                     & PIOLA                                              \\ 
\hline
8                     & MM2 CENTRALE                                       \\ 
\hline
9                     & CARACCIOLO                                         \\
\hline
\end{tabular}

}
        	\caption{$N_u$}
        	\label{fig:taxomony_example_n_u}
	    \end{subfigure}
	    \begin{subfigure}{.48\textwidth}
        	\centering
        	{
\tiny
\centering
\begin{tabular}{|l|l|} 
\cline{2-2}
\multicolumn{1}{l|}{} & \multicolumn{1}{c|}{\textbf{Location area}}        \\ 
\hline
1                     & ABRUZZI  \\ 
\hline
2                     & ACCURSIO                                           \\ 
\hline
3                    & LAGOSTA                                            \\ 
\hline
4                     & NIGUARDA                                           \\ 
\hline
5                     & PORTA NUOVA                                        \\
\hline
\end{tabular}

}
        	\caption{$SL = N_u \cap \widehat{T_u}$}
        	\label{fig:taxomony_example_sl}
	    \end{subfigure}
	     \begin{subfigure}{.48\textwidth}
        	\centering
        	{
\tiny
\centering
\begin{tabular}{|l|l|} 
\cline{2-2}
\multicolumn{1}{l|}{} & \multicolumn{1}{c|}{\textbf{Location area}}        \\ 
\hline
1                     & PIOLA                                         \\ 
\hline
2                     & MM2 CENTRALE                                             \\ 
\hline
3                     & CARACCIOLO                                             \\ 
\hline
4                     & BACONE 
            \\   
\hline
\end{tabular}

}
        	\caption{$TL=N_u\setminus \widehat{T_u}$}
        	\label{fig:taxomony_example_tl}
	    \end{subfigure}
	     \begin{subfigure}{1\textwidth}
        	\centering
        	{
\tiny
\centering
\begin{tabular}{|l|l|} 
\cline{2-2}
\multicolumn{1}{l|}{} & \multicolumn{1}{c|}{\textbf{Location area}}        \\ 
\hline
1                     & BELGIOIOSO                                         \\ 
\hline
2                     & CHIESE                                             \\ 
\hline
3                     & GORINI                                             \\ 
\hline
4                    & MINCIO                                             \\ 
\hline
\end{tabular}

}
        	\caption{$PL = \widehat{T_u}\setminus N_u$}
        	\label{fig:taxomony_example_pl}
	    \end{subfigure}
	    \begin{subfigure}{1\textwidth}
        	\centering
        	\vspace{5mm}
        	{
\footnotesize
\centering
\begin{tabular}{|c|c|} 
\hline
\textbf{Transit locations}      & \textbf{Significant locations}\\                     
\%TL = $\frac{|TL|}{|T_u|}$      & \%SL = $\frac{|SL|}{|T_u|}$   \\ 
5.6\%                           & 6.9\%                         \\
\hline
\textbf{Insignificant locations}& \textbf{Sporadic locations}\\                     
\%IL = $\frac{|IL|}{|T_u|}$      & \%PL = $\frac{|PL|}{|T_u|}$   \\ 
81.9\%                          & 5.6\%                         \\
\hline
\end{tabular}
}
        	\caption{Percentage of locations in classes}
        	\label{fig:taxomony_example_perc}
	    \end{subfigure}
	    
	\end{subfigure}
	\caption{Partitioning  of locations in classes,  based on the proposed taxonomy,  for the trajectory with ID=1470401.  The set of insignificant locations is omitted for readability. (a) The set $T_u$ of distinct locations in the native trajectory ordered by frequency descending order. (b) The set $\widehat{T_u}$ of distinct locations in the summary trajectory. (c) The set $N_u$ of the top frequent locations. (d) to (f): the location taxonomy; (d) the set of locations in the significant class $SL$, (e) the set of locations in the transit class $TL$, (f) the set of locations in the sporadic class $PL$. (g) Percentage of locations in each class}
	
	\label{fig:taxomony_example}
\end{figure*}


\section{Location diversity analysis: metrics}
In this section we extend the analysis from locations to trajectories and discuss possible metrics for measuring the individual mobility, focusing in particular on entropy-based metrics. 
We start presenting some background on diversity indices,  hence we introduce  the  metrics of concern.

\subsection{Diversity indices}
\subsubsection{Richness, Shannon-Weiner, Simpson}

We briefly overview three among the most popular diversity indices: richness, Shannon-Wiener, and Simpson. To illustrate the differences, consider an example population of 99 elements of type $a$,  and 1 element of type $b$.
\begin{itemize}
\item \textbf{Richness.} Richness (R) is given by the number of types \cite{2006oikos}. 
This metric does not take into account occurrences (or type \emph{abundance}).  Therefore, the population of the running example presents the same degree of diversity (R=2) of a population in which the elements of type $a$ and type $b$ are equally common. This index can thus yield too coarse results.

\item \textbf{Shannon-Wiener index.}
The Shannon-Wiener index is sensitive to the relative frequency of types.
The index is based on the Shannon entropy: $$H=-\sum_{i=1}^R p_i \ln p_i$$ where $p_i$ is 
the percentage of elements of type $i$, $R$ the richness (i.e. number of types), and $ln$ the natural logarithm. The entropy is maximal if types are equally common, i.e. $H_{max}=\ln{R}$, and minimal (i.e. H=0) if there is a unique type. In the previous  example, the entropy is: $H \approx 0.056$.  

\item \textbf{Simpson index.} This index is defined as: 
$$ S= \sum_{i=1}^R p_i^2$$
It holds that $S \geq \frac{1}{R}$, with $S=1/R$ when the types are equally common. With reference to the previous example, we have $S \approx 0.98$.
\end{itemize}

\subsubsection{True diversity} 
Diversity measures are hard to interpret and compare.  
For example, given a population $A$ with 8 equally common types and another population $B$ with 16 equally common types, the intuition is that the diversity of B is double the diversity of A. Yet, the Shannon entropy (calculated using base 2 logarithm) is 3 for $A$, and 4 for $B$. This is because entropy gives the uncertainty in type of a sample, not the number of types in the set \cite{2006oikos}.  
A different approach  
measures the diversity in terms of \emph{Hill numbers}. 
Proposed in the second half of the past century, Hill numbers
have been recently  reintroduced by biologists, e.g., \cite{2006oikos,2010tuo,2014chao}.
Hill numbers are a  family of diversity indices, differing among themselves only by an exponent $q$, and expressing diversity in terms of ''number of types''. The mathematical formulation is:
 
\begin{equation}
\label{eq:D}
^qD= (\sum_{i=1}^{R} p_i^q)^{\frac{1}{1-q}}
\end{equation} 

The measure $^qD$ is also called \emph{true diversity}, while $q \geq 0$ is the \emph{diversity order}. Popular diversity indices and true diversity  are tightly interrelated:
\begin{itemize}
\item  q=0:  true diversity  equals  richness, i.e. $^0D=R$
\item q=2: the true diversity is the inverse of the Simpson index, i.e.
\begin{equation}
^2D=1/(\sum_{i=1}^R p_i^2)
\end{equation}
\item q=1: Eq. \ref{eq:D} is not defined, but its limit exists and equals the exponential of the Shannon-Wiener entropy: 
\begin{equation}
^1D= \exp(-\sum_{i=1}^{R}p_i \ln{p_i})=\exp({H})
\end{equation}
\end{itemize}
While we refer to  the literature for further details, we  highlight a few  features that are relevant for our study:
\begin{itemize}
	\item 
The measures of popular  diversity indices can be translated into true diversity, and thus be expressed in terms of number of types. 
 Intuitively, given the value of a diversity index, say $x$, the corresponding true diversity indicates the number of equally common types in an ``equivalent'' population exhibiting  diversity $x$. 
\item The diversity order indicates the sensitivity of the diversity measure to the relative frequencies of types. 
In particular, the exponential of the Shannon-Wiener index (q=1) weights types by their frequency, without disproportionately favoring either
rare or common types.  In contrast, the inverse of the Simpson index (q=2)  places more weight on the frequencies of
abundant types and discounts rare types. In this sense it measures the diversity of the most popular types \cite{2014chao}. %

\end{itemize}

\subsection{True location diversity and location diversity profile}
We introduce the notion of true location diversity by extension. The set of locations in a summary trajectory represents the population of concern, where locations have a type and a number of occurrences.  
We consider the three indices: $R$, indicating the location richness;  the true location diversity of order 1, based on Shannon Entropy, denoted $TD_H$; and the true location diversity of order 2, the inverse of the Simpson index, denoted  $TD_S$.  
\noindent
\begin{example} Consider two trajectories $\widehat{T}_1$ and $\widehat{T}_2$. We only report, for brevity,  the series of locations forming the two trajectories, omitting the temporal information.  
\begin{itemize}
\item[]$\widehat{T}_1= a,b,c,d,e,a,b,c,d, e,a,b,c,d,e,a,b,c,d,e,a,b,c,d,e$
\item[]$\widehat{T}_2=a,b,a, b, a,b,a, c, a,b,a, b, a,b,a, d ,a, b, a, c, a,b,a, e, a$ 
\end{itemize}
Table \ref{tab:ex} shows the  relative frequencies of the 5 locations in the two trajectories.

\begin{table}[h]
	\centering
	\caption{Relative frequency of 5 types in $\widehat{T}_1$ and $\widehat{T}_2$}
	\begin{tabular}{l l l}
		\toprule
		Types & $\widehat{T}_1$ & $\widehat{T}_2$ \\
		\midrule
		a& 0.2 & 0.52  \\ 
    	b  & 0.2 & 0.32 \\ 
		c & 0.2 & 0.08 \\ 
		d  & 0.2 & 0.04\\ 
		e & 0.2 & 0.04\\
		\bottomrule
	\end{tabular}
	\label{tab:ex}
\end{table}
\begin{table}[h]
	\centering
	\caption{Diversity measures in $\widehat{T}_1$ and $\widehat{T}_2$}
	\begin{tabular}{l l l}
		\toprule
		Measures & $\widehat{T}_1$ & $\widehat{T}_2$ \\
		\midrule
		$R$& 5 & 5  \\ 
    	$TD_H$  & 5 & 3.2 \\  
		$TD_S$ & 5 & 2.6 \\
		\bottomrule
	\end{tabular}
	\label{tab:ex1}
\end{table}
The diversity measures, homogeneously expressed in term of 'number of types', are  reported in Table  \ref{tab:ex1}. 

It can be seen that the diversity of order $>0$ is significantly lower than $R=5$, coherently with the fact that locations are not evenly frequented.  The value $TD_H= 3.2$ measures the diversity of the typical locations in $\widehat{T}_2$, while $TD_S = 2.6$, the diversity of the highly frequented locations.
\end{example}

\noindent
\paragraph{Location diversity profile.} 
Diversity measures provide complementary information on mobility. Since all of them are relevant,
we follow the approach  presented by ecologists in \cite{2014chao}, and introduce the notion of \emph{location diversity profile}. The location diversity profile  provides a multi-level characterization of location heterogeneity through a multiplicity of diversity indicators, globally enhancing the interpretation of the user behavior. To better illustrate the information conveyed by the diversity profile, we present an example. 
\begin{Def}[Location diversity profile]
The location diversity profile of trajectory $i$ is defined by the triple:
\begin{equation}
Pr(i)=(R^i, TD_H^i, TD_S^i)
\end{equation}
\end{Def}

\noindent

\begin{figure}[h] 
	\centering
	\begin{subfigure}{0.485\textwidth}
	\centering
	    \includegraphics[width=.65\linewidth]{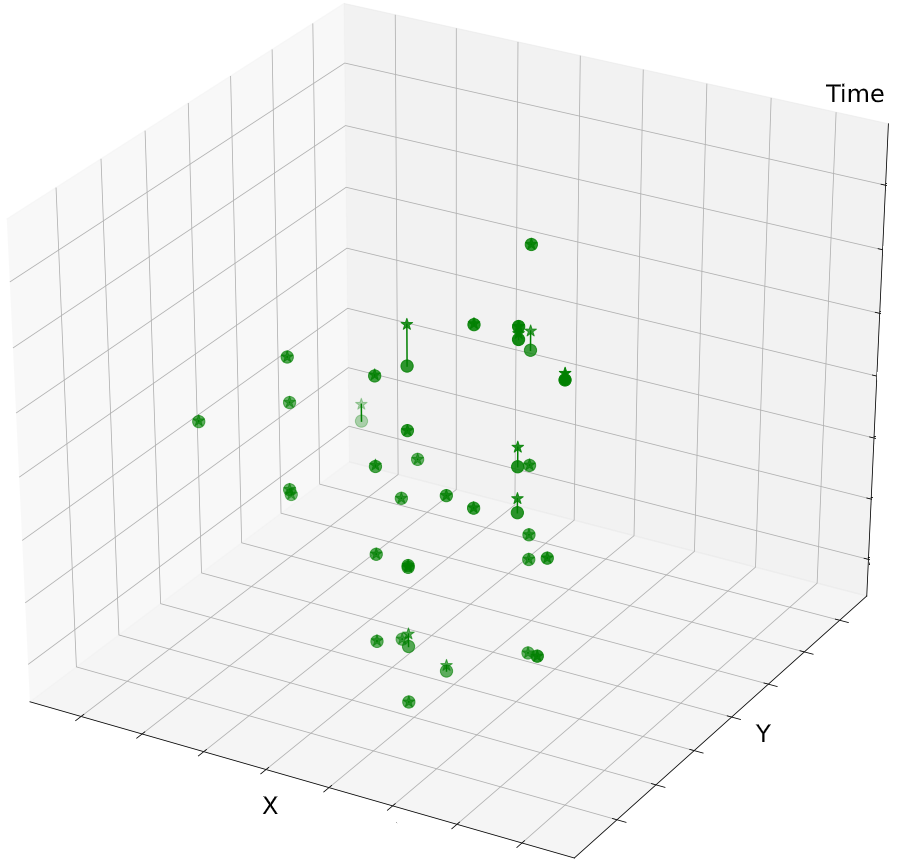}
	    \caption{ID 1595546}
	\end{subfigure}
	\begin{subfigure}{0.485\textwidth}
	\centering
	    \includegraphics[width=.65\linewidth]{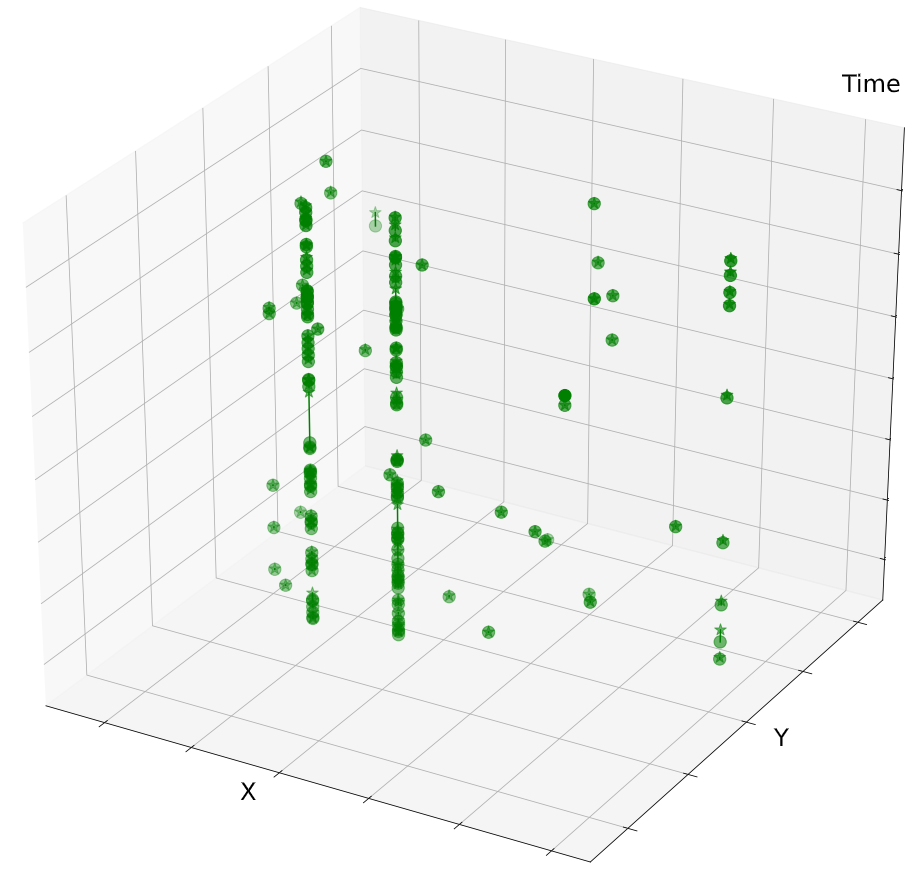}
	    \caption{ID 1686599}
	\end{subfigure}
	\begin{subfigure}{0.485\textwidth}
	\centering
	    \includegraphics[width=.65\linewidth]{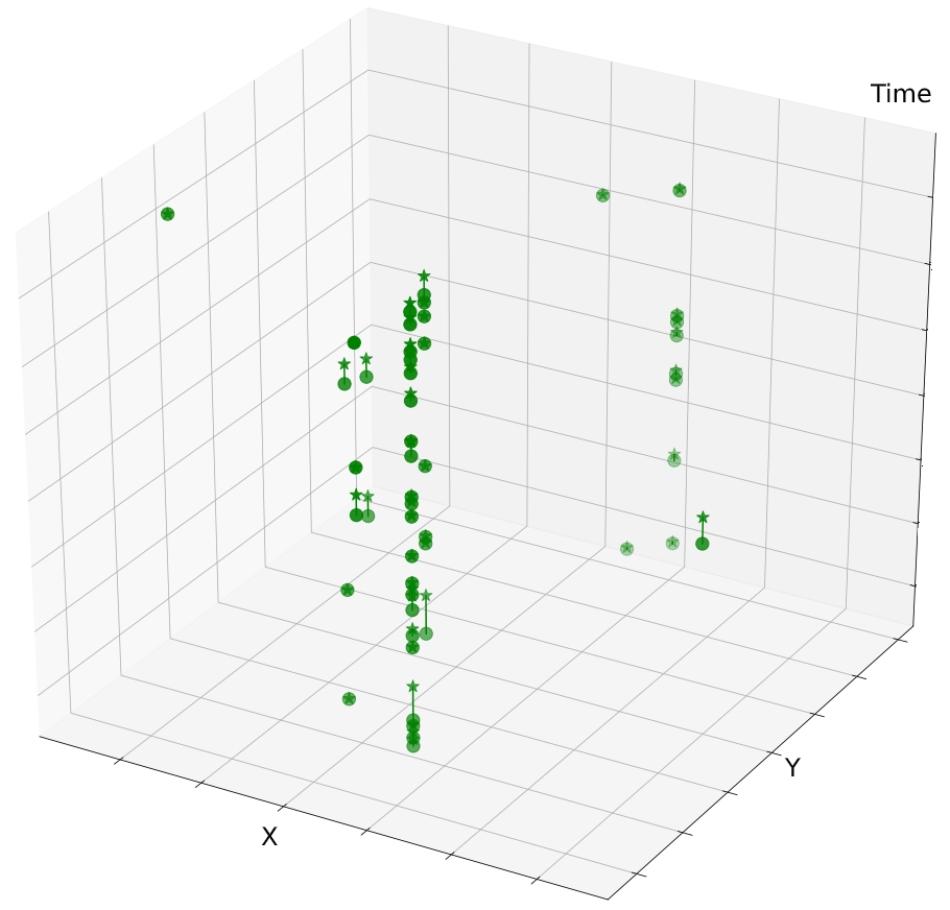}
	    \caption{ID 1477366}
	\end{subfigure}
	\begin{subfigure}{0.485\textwidth}
	\centering
	    \includegraphics[width=.65\linewidth]{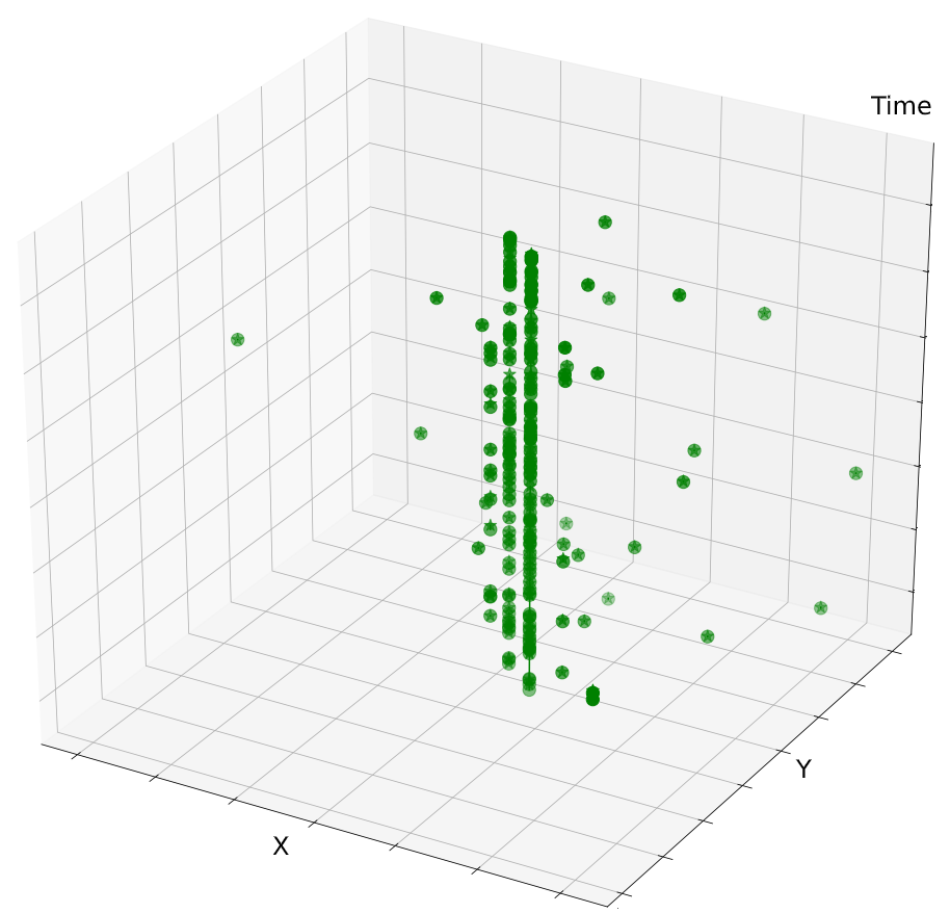}
	    \caption{ID 1733146}
	\end{subfigure}

	
	\caption{Summary trajectories represented in space (axes x,y) and time (z). The points in space are the centroids of the Location Areas }
	\label{entropy_vs_profile_example2}
\end{figure}

\noindent
\begin{example} 
 Table \ref{tab:entropy_vs_diversityprofile} reports, for comparison, the Shannon entropy and the location diversity profile of four summary  trajectories from the reference dataset.  The Table reports two cases, discussed below: 
 
 \begin{table}[h]
	\centering
	
	\caption{Shannon-Wiener entropy (H) and location diversity profile}
	\begin{tabular}{ c c c c c }
		\toprule
		 Traj. ID & $H$ & $R$ & $TD_H$ & $TD_S$\\
		\midrule
        1595546 & 2.48 & 15 & 12.01 & 9.97 \\
        1686599 & 1.96 & 30 & 7.07 & 3.8 \\
         1477366 & 1.74 & 10 & 5.73 & 4.03 \\
        1733146 & 1.74 & 30 & 5.74 & 3.31 \\
        \bottomrule

	\end{tabular}
	\label{tab:entropy_vs_diversityprofile}
\end{table}

\begin{itemize}
    \item
 (Case 1). Consider the  first two trajectories $\widehat{T_1}$=1595546 and $\widehat{T_2}$=1686599. We can see that entropy $H$ is greater in $\widehat{T_1}$. Thus, we can infer that the uncertainty in type of a location randomly selected in $\widehat{T_1}$ is greater that that of a location selected in $\widehat{T_2}$. The diversity profile, however, provides more explicit information. 
In fact, we can see that  $\widehat{T_1}$ contains fewer locations than $\widehat{T_2}$  (i.e. $15$ locations instead of $30$); in addition, the locations in $\widehat{T_1}$ are nearly all equally frequented.  The second user instead, although visiting  double the number of locations, frequent intensively only few locations.  
The behavior of the two users is thus quite different and that is also evident in the plots in Figure \ref{entropy_vs_profile_example2}.(a)-(b),  representing the summary trajectories in space and time at population scale;

\item
(Case 2). Consider now the last two trajectories $\widehat{T_3}$=1477366 and $\widehat{T_4}$=1733146. In this case, the value of $H$ is identical. Yet, the diversity profile shows that the two users frequent a different number of locations (i.e. $T_3$ reports 10 locations, $T_4$, 30 locations), while  the number of location types in $TD_H$ and $TD_S$  is similar. The two users thus exhibit a different behavior, and that finds visual confirmation in  the plots in Figure \ref{entropy_vs_profile_example2}.(c)-(d). 
\end{itemize}
\end{example}

\subsection{Entropy rate}

The diversity indices and derived measures presented in the previous section are all based on either the number of unique visited locations or their relative frequency, so that – for example – the two trajectories
\begin{itemize}
\item[]$\widehat{T}_1=a,a,a,b,b,c$
\item[]$\widehat{T}_2=a,b,a,c,a,b$ 
\end{itemize}
have the same location diversity profile.
Clearly these two trajectories convey different information about the mobility of the user they refer to;
in other words, intuition suggests that, in addition to the frequency of visitation, the \emph{order} in which locations are visited may also be a relevant property in the assessment of the mobility behavior of an individual.

To this intent, we consider the \emph{entropy rate} as another useful indicator of users' mobility.
Entropy rate is a well-known indicator and has already been used in \cite{2010barabasiPredictability} to estimate an upper bound of predictability in human mobility.
In the following, we summarize its definition and propose an algorithm to estimate the entropy rate from finite sequences.

\subsubsection{Entropy rate of stochastic processes}

The content presented in this subsection is derived from \cite{ElementsOfInformationTheory}, to which the reader is invited to refer to for a more detailed explanation.

Let $\mathcal{X}=(X_i)_1^{+\infty}$ be a discrete-valued stochastic process.
The \emph{entropy rate} of $\mathcal{X}$ is defined as
\begin{equation}
\label{eq: entropy rate}
H_r(\mathcal{X})=\lim_{n\to+\infty}\frac{1}{n}H(X_1,X_2,\ldots,X_n)
\end{equation}
(when the limit exists) where $H$ denotes entropy.
This definition expresses the notion that entropy rate is the per-symbol entropy of the $n$ jointly distributed random variables $X_1,X_2,\ldots,X_n$.

If the process $\mathcal{X}$ is \emph{stationary} (i.e. the joint distribution of any subset of random variables is invariant with respect to shifts in the index), it can be shown that $H_r(\mathcal{X})$ exists and equals to
\begin{equation}
\label{eq: entropy rate stationary}
H_r(\mathcal{X})=\lim_{n\to+\infty}H(X_n\mid X_1,X_2,\ldots,X_{n-1}),
\end{equation}
thus expressing that, in this case, the entropy rate is the conditional entropy of the last random variable given the past.

\subsubsection{Entropy rate estimation for finite sequences}

Let $\mathbf{x}=(x_1,x_2,\ldots,x_n)$ be a data realization of length $n$ drawn from the stochastic process $\mathcal{X}$.
Given only $\mathbf{x}$, an exact value for the entropy rate of the originating process $\mathcal{X}$ can not be calculated, not only because we observe a merely finite-length sample of the process, so we can not compute the limit for $n\to+\infty$, but also because the joint probability distributions (of $X_1,\ldots,X_n$ in \eqref{eq: entropy rate} and of $X_1,\ldots,X_{n-1}$ in \eqref{eq: entropy rate stationary}) are not known.
To be able to calculate any statistical estimate, a necessary assumption is that the process is \emph{ergodic}, i.e. that its statistical properties can be deduced from a sufficiently long random sample.

Several entropy rate estimators have been proposed with varying requirements, theoretical properties and performance.
Here we describe the estimator defined by Theorem 1 c) in \cite{1998Kontoyiannis}, derived from Lempel-Ziv compression algorithm.
This estimator has a good combination of desirable properties:
it is consistent (i.e. for $n\to+\infty$ it converges to the entropy rate of the process);
it does not require preprocessing of data;
is parameter-free;
it has good performance on short sequences;
is very easy to implement.
The only hypothesis on $\mathcal{X}$ needed by this estimator, other than the ergodicity of the process, is the so-called \emph{Doeblin Condition}, which requires that, after some number of steps, every outcome is possible again with positive probability, independently of whatever may have occurred in the past (see \cite{1998Kontoyiannis} for a formal definition).

Given a sequence $\mathbf{x}=(x_1,x_2,\ldots,x_n)$ of length $n$ and the current position $i\in\{1,2,\ldots,n\}$ in the sequence, the main quantity of interest of the presented estimator is the longest match-length
\begin{equation}
l_i(\mathbf{x}) = \max\left\{k\;\middle\vert\;
\begin{aligned}
&k\in\{0,\ldots,n-i+1\}\\
&(x_i,\ldots,x_{i+k-1})=(x_j,\ldots,x_{j+k-1})\\
&j\in\{1,\ldots,i-1\}
\end{aligned}
\right\},
\end{equation}
that is the length of the longest subsequence starting from $i$ that has already been encountered before.
The quantity
\begin{equation}
\label{eq: entropy rate estimate}
\widetilde{H}_r(\mathbf{x})=\frac{n\log_2(n)}{\sum_{i=1}^n(l_i(\mathbf{x})+1)}
\end{equation}
is an estimator for $H_r(\mathcal{X})$.
To compute this estimate we propose a straightforward algorithm \footnote{A Julia implementation of the proposed algorithm is publicly available at \url{https://gist.github.com/matteorossini/1c53b28efad63a8326e418bd1c3645a2}.} (see Algorithm~\ref{alg: entropy rate estimation}) having an $O(n^3)$ worst-case complexity.

\begin{algorithm}[h]
\caption{Entropy rate estimation.}
\label{alg: entropy rate estimation}
\DontPrintSemicolon
\SetKwData{i}{i}
\SetKwData{j}{j}
\SetKwData{k}{k}
\SetKwData{l}{l}
\SetKwData{n}{n}
\SetKwData{sum}{sum}
\SetKwFunction{length}{length}

\hrule
\vspace*{0.3cm}
\KwData{a finite sequence $(x_1,\ldots,x_n)$}
\KwResult{an estimate of the entropy rate of the process from which $(x_1,\ldots,x_n)$ is drawn}
\n $\leftarrow$ \length($(x_1,\ldots,x_n)$)\;
\If{$\n\leq1$}{
    \Return{$0$}
}
$\sum\leftarrow\n$\;
\For{$\i\leftarrow2$ \KwTo \n}{
    \l $\leftarrow$ 0\;
    \For{$\j\leftarrow1$ \KwTo $\i-1$}{
        \k $\leftarrow$ 0\;
        \While(\tcp*[h]{$\&\&$: short-circuit AND}){$\i+\k\leq\n\;\&\&\;x_{\j+\k}==x_{\i+\k}$}{
            $\k\leftarrow\k+1$
        }
        \If{$\k>\l$}{
            $\l\leftarrow\k$
        }
    }
    $\sum\leftarrow\sum+\l$
}
\Return{$\frac{\n\log_2(\n)}{\sum}$}
\vspace*{0.3cm}
\hrule

\end{algorithm}

\subsubsection{Entropy rate estimation for trajectories}

A summary trajectory can be considered as a finite-length sample of a discrete-valued time-correlated ergodic process; the Doeblin Condition is clearly satisfied in our application, where the history of locations visited by a user does not limit the location that they could visit in the future, so the entropy rate can be estimated from trajectories using \eqref{eq: entropy rate estimate}.
\begin{figure}[h]
    \centering
    \begin{subfigure}{0.485\textwidth}
        \centering
        \includegraphics[width=0.7\textwidth]{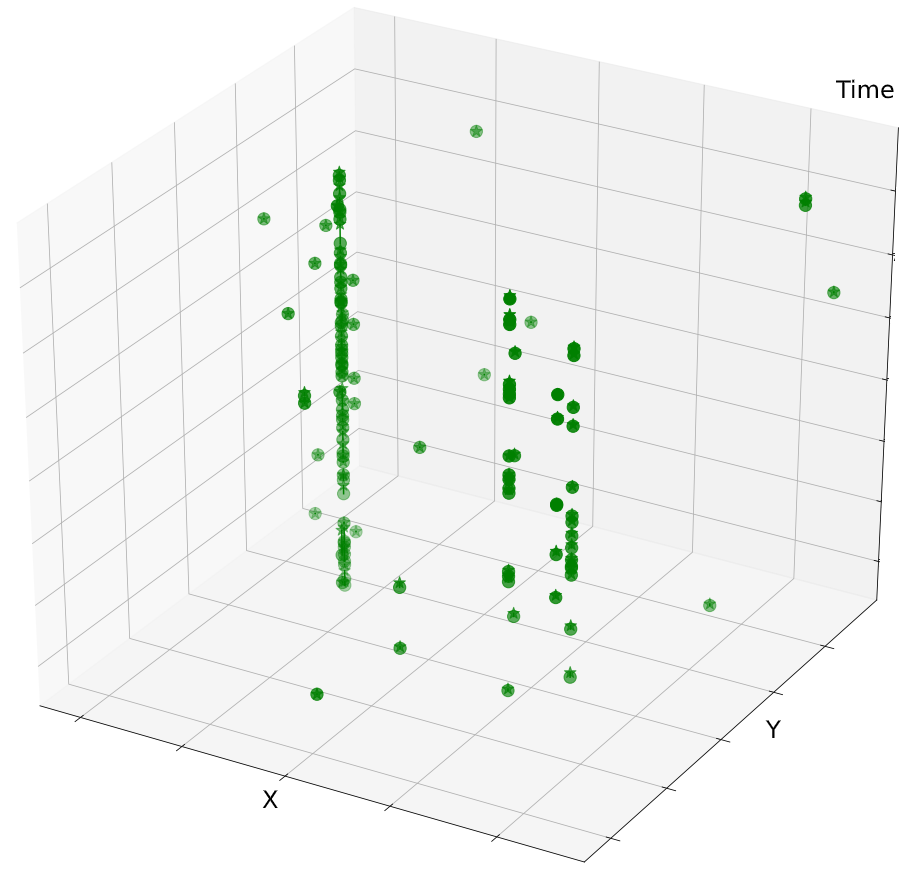}
        \caption{}
    \end{subfigure}
    \begin{subfigure}{0.485\textwidth}
        \centering
        \includegraphics[width=0.7\textwidth]{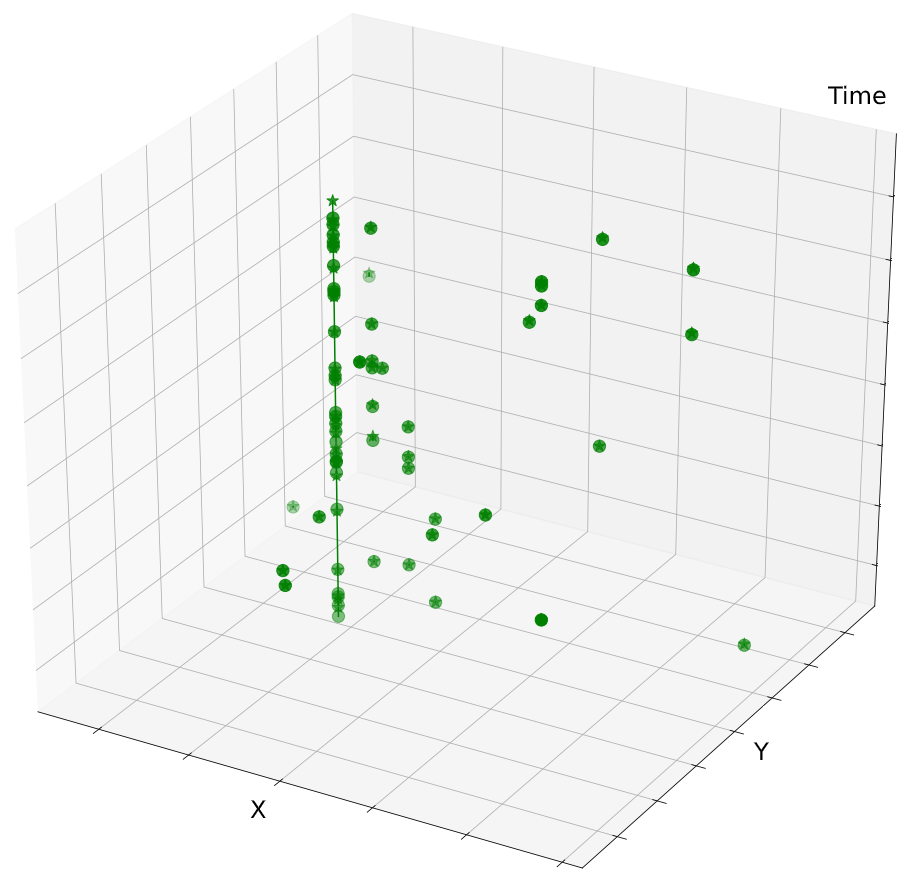}
        \caption{}
    \end{subfigure}
    \caption{
       Summary trajectories  in space and time:
        (a) Trajectory 1748734. (b) Trajectory 1900971.
        Despite having very similar location diversity profile, namely $Pr(1748734)=(16,7.34,3.88)$ and $Pr(1900971)=(16,7.31,3.87)$, the entropy rates of these two users are quite different (1.53 and 3.12 bits respectively) suggesting that their mobility behavior is different, as shown in the pictures.
        }
    \label{fig: entropy rate vs. location diversity profile}
\end{figure}
Note that, since the entropy rate depends on the order of visited locations, it carries different information with respect to the indicators discussed in the previous section.
Therefore the entropy rate can be useful to distinguish users who have different mobility patterns despite having the same location diversity profile.
An example from the reference dataset is reported in Figure~\ref{fig: entropy rate vs. location diversity profile}.

%


\section{Experiments} 
We turn to evaluate the analytical framework through a series of  experiments conducted on the  trajectory dataset described in Section \ref{data}. 
The experimental activity is split into the following four tasks:
\begin{itemize}
    \item To evaluate the SeqScan-d algorithm against the RLE-centric baseline method. The goal is to compare the summarization capabilities;
  \item To validate the hypothesis that attractiveness and frequency are two  distinct though interrelated dimensions of location relevance. Definitely, the goal is to quantify the relation between frequent locations in native trajectories and attractive locations in summary trajectories, at population scale;
  \item To evaluate the utility of summary trajectories. The goal is to assess whether major statistical properties of telco trajectories reported in the literature are preserved when trajectories are summarized;
  \item To compute diversity measures and related metrics at population scale, and  relate, where possible, those results to the literature on human mobility modeling.   
  
\end{itemize}
In the following, we  describe  the experiments conducted to accomplish the above tasks. The experimental setting is defined below.

\subsection{Experiment setting}
\paragraph{Dataset.}  
We recall that the dataset  consists of 100,000+ telco trajectories in the area of Milan, of various length and duration, over a period of  67 days. The total number of samples in the dataset  amounts to about 54 million points.  The \emph{telco space} consists of 685 locations at the granularity of Location Area, and  identified by a label. Table \ref{tab:statsdb} reports the summary statistics on the number of trajectories (i.e. users), number of records,  average  and standard deviation of trajectory length. 
\begin{table}[h]
	\centering 

	\caption{Summary statistics of the dataset }	
	\begin{tabular}{c c c c c}
		\toprule
		\# Traj  & \# Records & \# Loc   & Avg(trj\_len)  & Std(trj\_len) \\
		\midrule
		104,413 &54,193,257& 685& 3151  & 1650\\ 
		\bottomrule
	\end{tabular}	
	\label{tab:statsdb}
\end{table}

\noindent
\paragraph{Hw/sw platform.} Data summarization is performed on a Linux-Ubuntu Server DELL T620, 362 GB Ram, equipped with 2 six-cores XEON CPUs that provide 24 logic CPUs.  The summarization operation is performed by partitioning the input dataset among the CPUs, and processing in parallel the smaller datasets.  Data analysis is  performed on a standard Windows 10 PC.

\subsection{Evaluation of the summarization technique}
We conduct three experiments: (a) the first is to analyze the sensitivity of summarization to the SeqScan-d parameters. This also provides the basis for the choice of the parameter configurations used in the rest of this Section. (b) The second experiment details the reduction in the number of distinct locations due to summarization. (c) The third and core experiment compares the summary trajectories generated by SeqScan-d  with the set of RLE+ trajectories.

\subsubsection{Parameter sensitivity analysis}
We start running the summarization algorithm 
with different values for the clustering parameters $N$ and $\delta$.
Table \ref{tab:par} reports the chosen parameters with $N$ ranging in $\{2,4,6,8\}$ and $\delta$ varying in 8 intervals from $0'$ to $120'$. We run the algorithm for each combination of the input values, obtaining  32 summarized datasets $\{\widehat{D}_1,..,\widehat{D}_{32}\}$. Hence, for each  summary dataset $\widehat{D}_i$, we compute  the two performance measures, the summarization rate $S_{rate}(\widehat{D}_i)$ and the summarization goodness $Q(\widehat{D}_i)$.

\begin{figure*}[t] 
		\centering
		\begin{subfigure}[b]{.485\linewidth}
            \includegraphics[width=\textwidth]{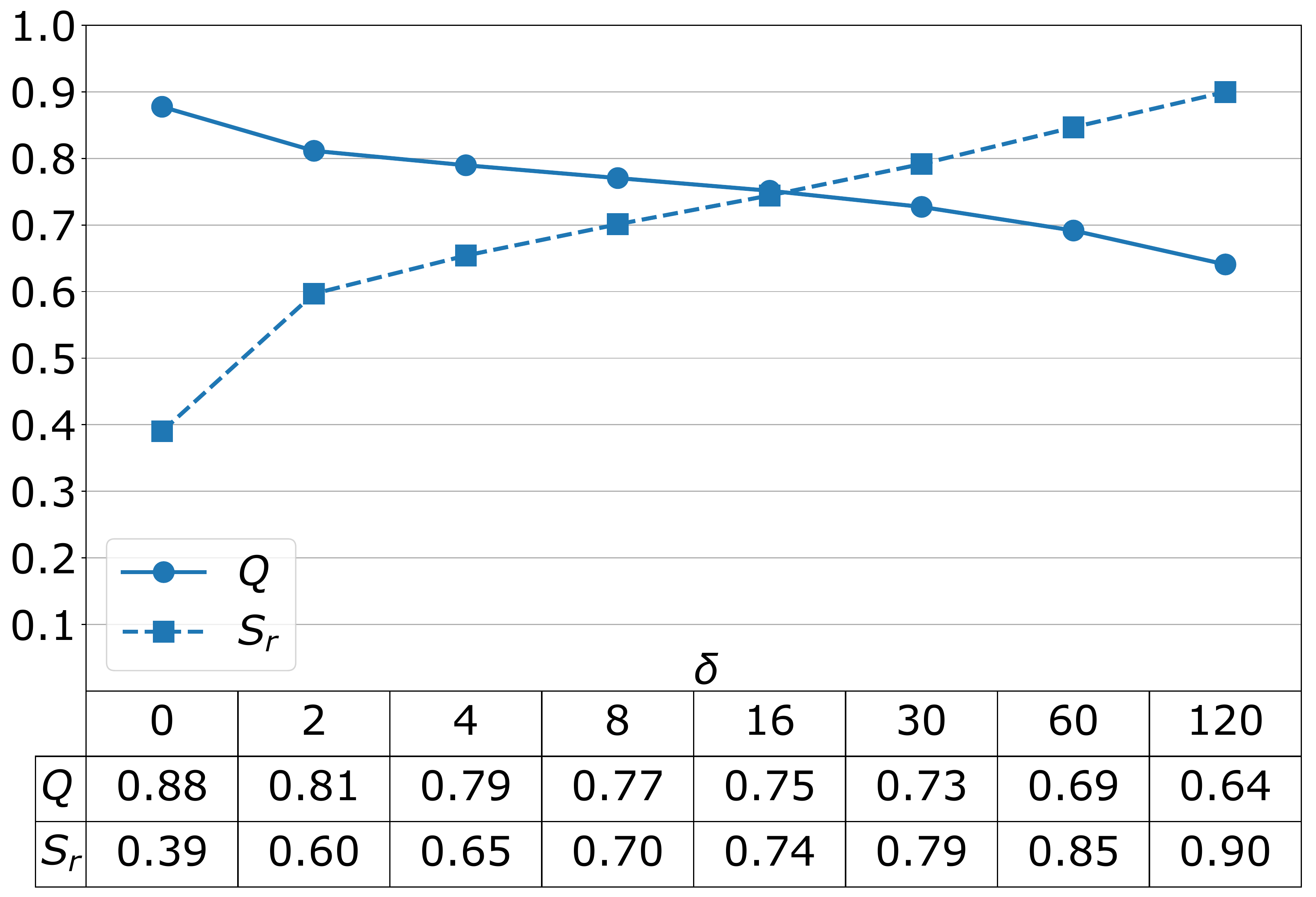}
            \caption{$N$=2}
            \label{AvgQ_S_many_params_n2}
        \end{subfigure}
        \begin{subfigure}[b]{.485\linewidth}
            \includegraphics[width=\textwidth]{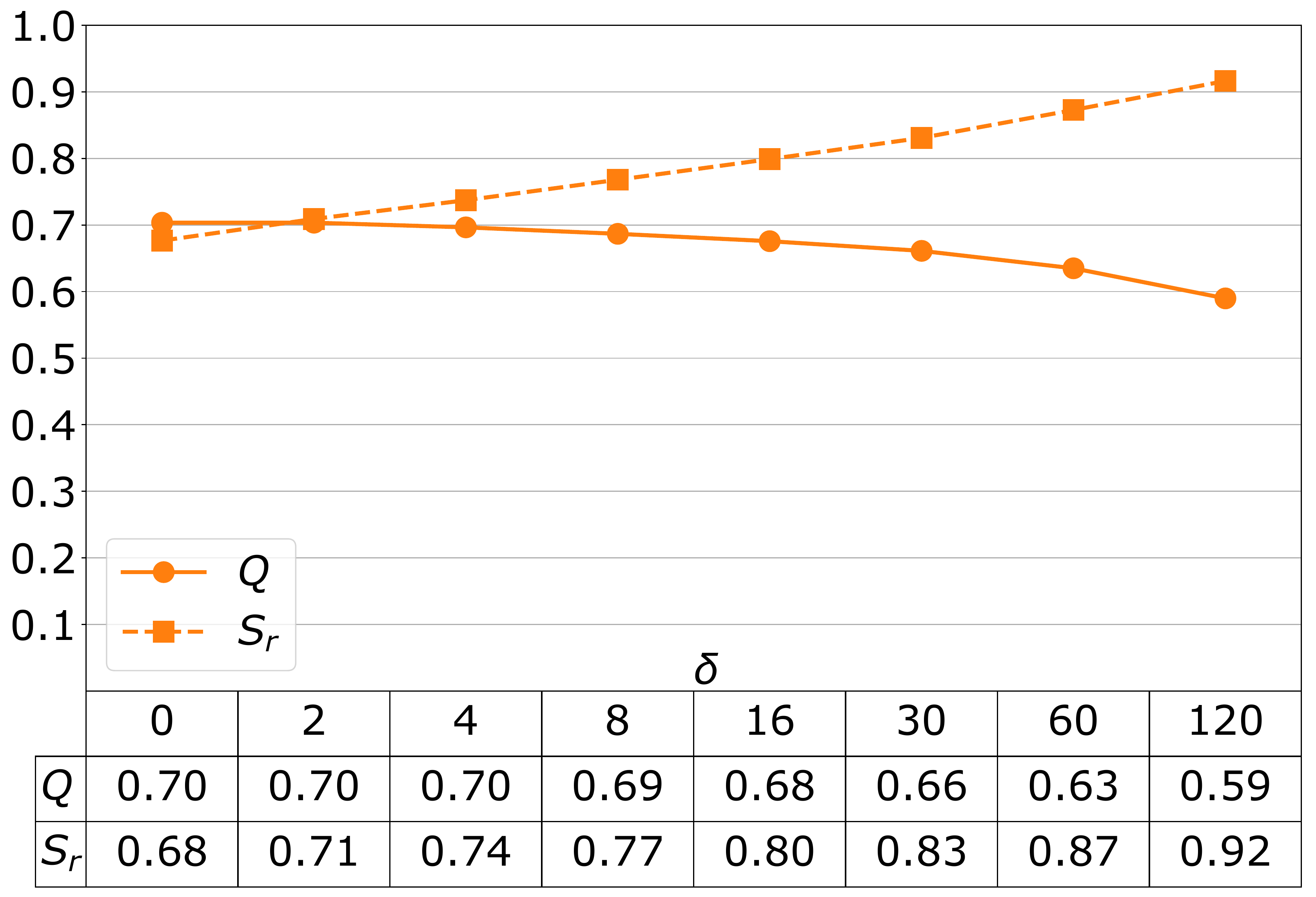}
            \caption{$N$=4}
            \label{AvgQ_S_many_params_n4}
        \end{subfigure}
        \begin{subfigure}[b]{.485\linewidth}
            \includegraphics[width=\textwidth]{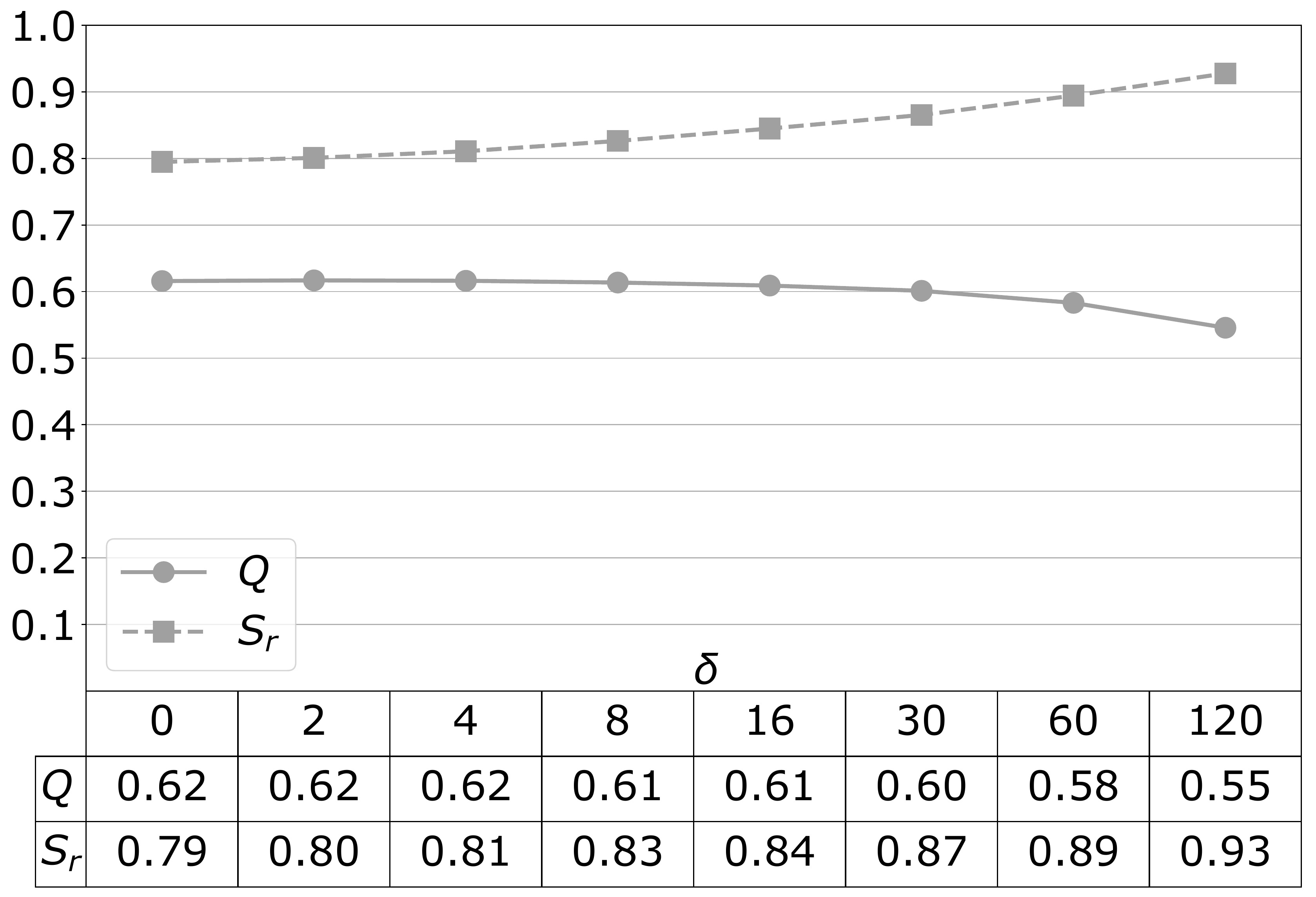}
            \caption{$N$=6}
            \label{AvgQ_S_many_params_n6}
        \end{subfigure}
        \begin{subfigure}[b]{.485\linewidth}
            \includegraphics[width=\textwidth]{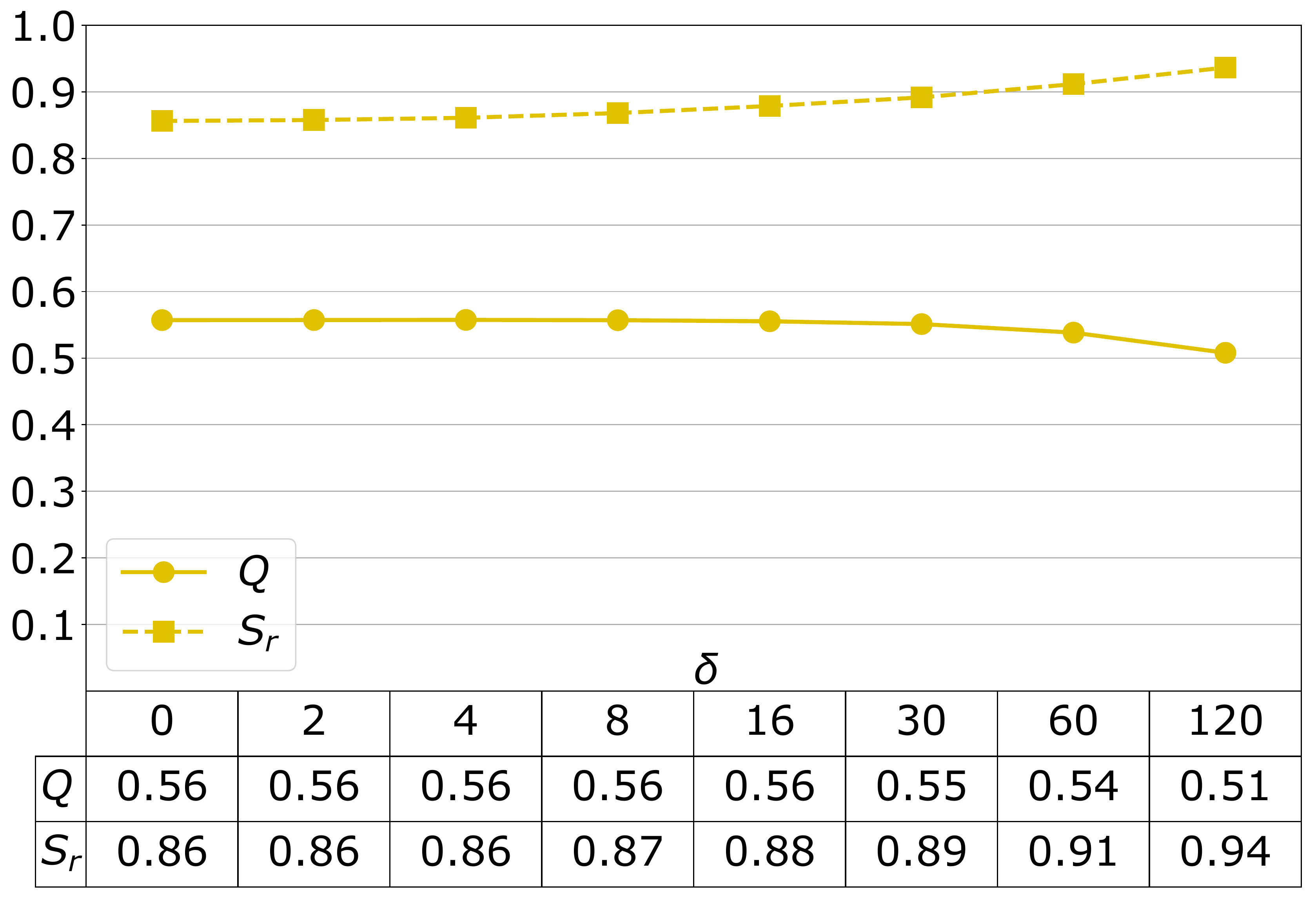}
            \caption{$N$=8}
            \label{AvgQ_S_many_params_n8}
        \end{subfigure}
		\caption{The summarization goodness $Q(D)$ and the summarization rate $Sr(D)$ for $\delta$ ranging in $\{0,2,4,8,16, 30, 60, 120\}$, and for $N$= 2, 4, 6 and 8, respectively}
	\label{AvgQ_S_many_params}
\end{figure*}

\begin{table}[h]
	\centering
	
	\caption{Input parameters for data summarization}
	\begin{tabular}{ c c }
		\toprule
		N & $\delta (day)$ \\
		\midrule
		2, 4, 6, 8 & 0.0000 $\approx 0'$ \\ 
		2, 4, 6, 8  & 0.0014 $\approx 2'$ \\ 
		2, 4, 6, 8 & 0.0028 $\approx 4'$ \\ 
		2, 4, 6, 8  & 0.0055 $\approx 8'$\\ 
		2, 4, 6, 8  & 0.0111 $\approx 16'$ \\  
		2, 4, 6, 8  & 0.0208 $\approx 30'$ \\ 
		2, 4, 6, 8  & 0.0417 $\approx 60'$\\ 
		2, 4, 6, 8  & 0.0833 $\approx 120'$ \\
		\bottomrule
	\end{tabular}
	\label{tab:par}
\end{table}
\noindent
 Figure \ref{AvgQ_S_many_params} shows the four plots reporting the performance measures for varying values of $\delta$, and $N=2,4,6,8$. 
\noindent
It can be seen that  
the goodness of summarization is generally high. For example, the plot in Figure \ref{AvgQ_S_many_params_n2} (for N=2) shows that, for $\delta=16'$, the measure is  0.75. Following the experience in \cite{Damiani2016}, the value of 0.7 is chosen as lower bound for a  clustering of high quality. 
It can be also seen that the summarization rate is quite high, with the percentage of irrelevant locations (types) varying approx from 40\% to 90\%. In particular, with $N=2$ and $\delta=16'$,  we achieve a summarization rate of 0.74.

The parameter $N$ has a strong impact on the summarization performance. That is not surprising:
the higher  $N$, the longer the sequences representing clusters and thus the chance of absences in the period. Therefore,  the highest goodness  is  achieved with $N=2$. 
In contrast, $\delta$ is an application-oriented parameter specifying the maximum temporal granularity of relevant locations. For example, the value $\delta=16'$ means that the temporal extent of a cluster shall be at least 16' (ruling out absences). 
For $N=2$, the two curves of the performance metrics intersect for $\delta=16'$, which is also a reasonable value, for example, when colocated events have to be detected \cite{Pedreschi_KDD_11}. Thus, for $N=2$ it is convenient to set $\delta =16'$.
As $N$ increases, 
both $Q(D)$ and $S_{rate}(D)$ vary slightly with its variation; even, for $N=8$ they are almost constant. Based on above considerations and for the sake of readability, in the following sections we will report the results only for the four series of trajectories obtained with N=2,4,6,8 and $\delta=16'$.

\begin{figure*}[t]
\begin{minipage}{\linewidth}
    \centering
    \begin{subfigure}[b]{.485\linewidth}
        \includegraphics[width=\textwidth]{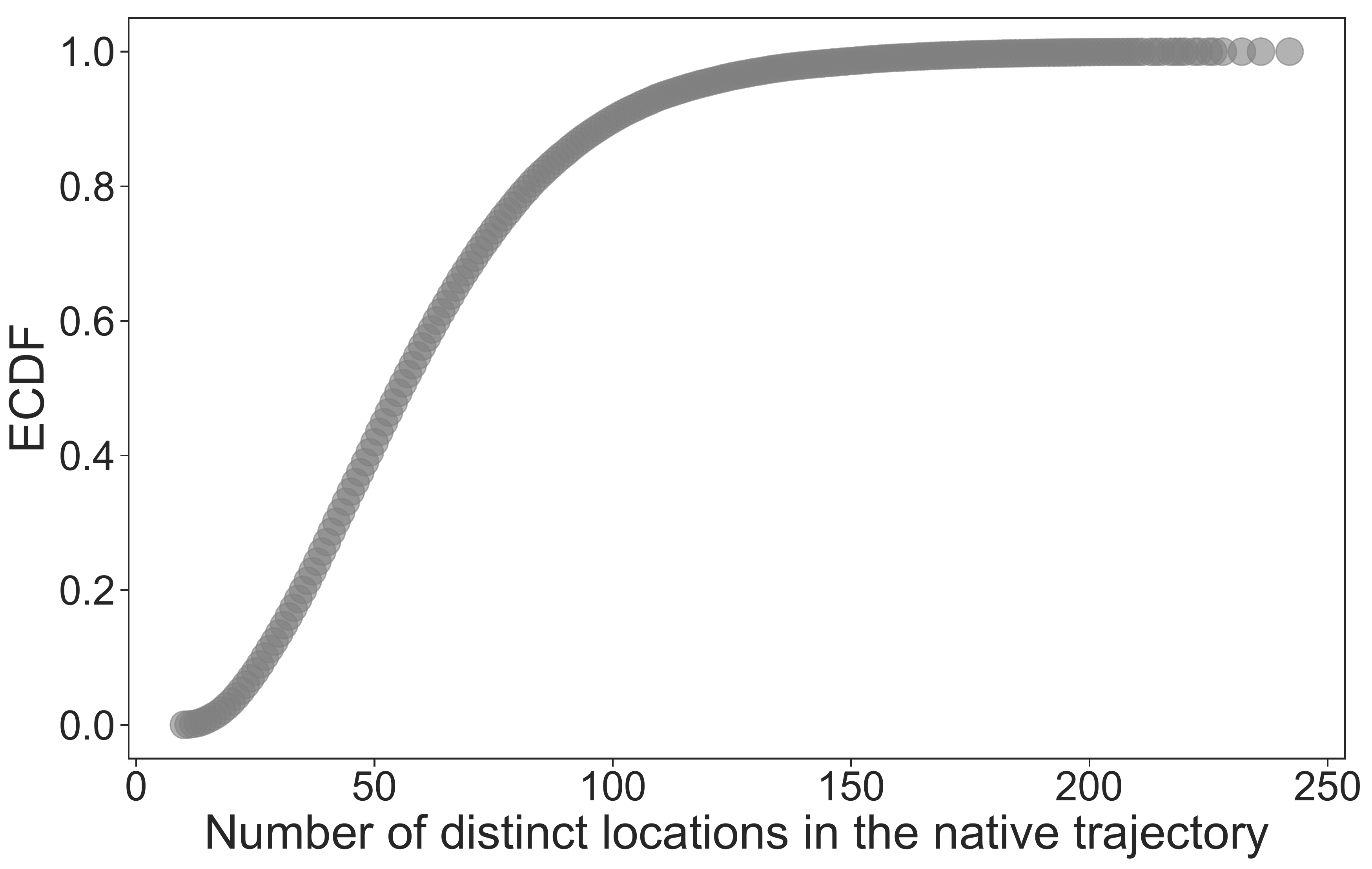}
        \caption{}\label{fig:cdf_num_location_orig}
    \end{subfigure}
    \begin{subfigure}[b]{.485\linewidth}
        \includegraphics[width=\textwidth]{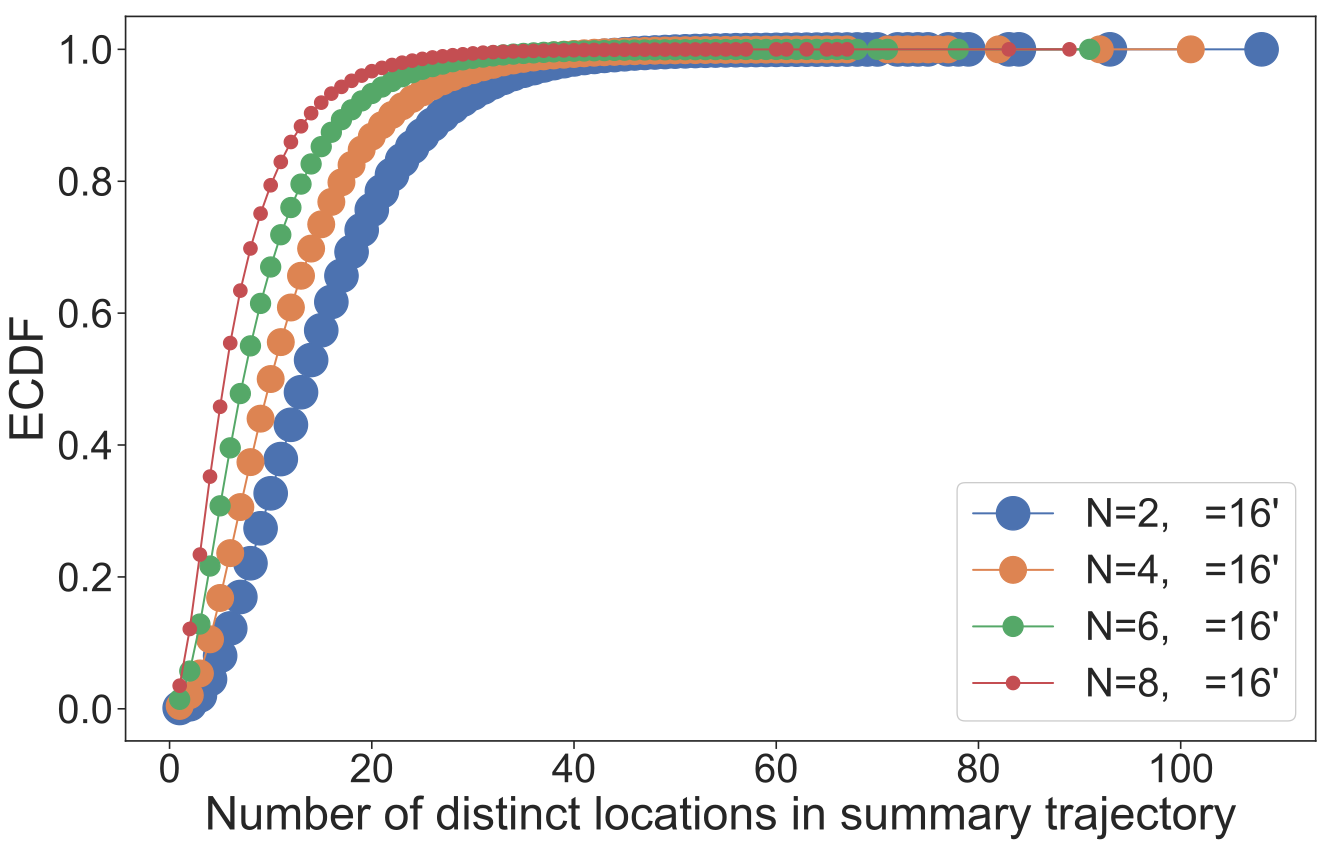}
        \caption{}\label{fig:cdf_num_location_cluster}
    \end{subfigure}

     \begin{subfigure}[b]{.485\linewidth}
    \vspace{3mm}
    \centering
    \begin{tabular}{ c c c c }
        \toprule
         Parameters & Median &  Mean &  Std \\
         \midrule
         Original & 56.0 & 61.09 & 30.11 \\
         N=2, $\delta$=16$'$ & 14.0 & 15.61  & 8.71 \\
         N=4, $\delta$=16$'$ & 10.0 & 12.28 &  7.75 \\
         N=6, $\delta$=16$'$ & 8.0 & 9.47 & 6.59 \\
         N=8, $\delta$=16$'$ & 6.0 & 7.40 & 5.57 \\
         \bottomrule
    \end{tabular}
    \caption{}\label{tab:num_location}
    \end{subfigure}
\end{minipage}
\caption{(a) Empirical Cumulative Distribution Function (ECDF) of the number of distinct locations in original trajectories. (b) ECDF of the number of distinct locations in summary trajectories. (c) Summary statistics of the number of visited  distinct locations in original trajectories and in summary trajectories}
    \label{fig:num_location}
\end{figure*}

\subsubsection{Number of distinct locations in summary trajectories}
This experiment provides further details on the set of locations obtained from summarization. Such a set is the basis of additional analysis that will be presented later on in the article.  \figurename~\ref{fig:num_location} shows the distribution of the cardinality of the location sets obtained by running SeqScan-d with the aforementioned parameter configurations. As an example, for $N=4$, summary trajectories contain on average 12 distinct locations against 61 location in the original trajectories. Such locations are those that, based on our model, are  \emph{attractive}.    


\subsubsection{Comparison of SeqScan-d with the RLE-centric baseline technique}
We compare the trajectories generated by SeqScan-d with 
the RLE+ trajectories. We recall that RLE+ segments denote sequences of the original trajectories reporting a unique location, thus RLE+ is  sensitive to noise. We hypothesize  that noise determines an excessive fragmentation of the input trajectories causing location information loss.  In this experiment, we want to evaluate the impact of noise on summarization capabilities, based on the following two metrics: (a) the number of distinct locations in resulting trajectories; (b) the trajectory length, i.e. the number of segments in the trajectory. Note that we do not refer to the quality indicator $Q$ (Section 4.2), because it makes sense only for SeqScan-d and, more in general, for noise insensitive techniques. 



Table \ref{tab:comp} reports the summary statistics about the two measures computed for the reference dataset. It can be seen that: 
\begin{itemize}
 \item  SeqScan-d generates shorter trajectories. For example, with $N=4, \delta=16'$, summary trajectories are around 33\% shorter than RLE+ trajectories. This result can be explained as follows: SeqScan-d not only can recognize RLE+ segments as key locations, but also is  capable of aggregating  RLE+ segments separated by noise, in a unique segment. The result is a reduced fragmentation. 
 \item SeqScan-d identifies a higher number of distinct locations. For example, with the above parameters, the average number of distinct locations in summary trajectories and RLE+ trajectories is 12.28 and 11.43, respectively. In general, the  set of locations types in the summary trajectories is a superset of the corresponding set in RLE+ trajectories. Thus the location information loss is lower.

\end{itemize}
  
  \begin{table*}[t]
	\centering
	\scriptsize
	\caption{RLE+ Vs. SeqScan-d. Summary statistics of the number of distinct locations per trajectory and trajectory length}
\begin{adjustbox}{width=\columnwidth,center}
\begin{tabular}{c c c c c c c c c c c c c }
\toprule
& \multicolumn{6}{c}{\textbf{Distinct locations}} & \multicolumn{6}{c}{\textbf{Trajectory length}}\\
\midrule
 &\multicolumn{3}{c}{\textbf{RLE+}} & \multicolumn{3}{c}{\textbf{SeqScan-d}} & \multicolumn{3}{c}{\textbf{RLE+}} & \multicolumn{3}{c}{\textbf{SeqScan-d}} \\
 \midrule
& Median & Mean & Std & Median & Mean & Std & Median & Mean & Std & Median & Mean & Std \\
\midrule
N=2 & 14 & 15.53 & 8.8 & 14 & 15.61 & 8.87 & 113 & 122.98 & 67.44 & 75 & 82.63 & 52.52 \\
N=4 & 10 & 11.43 & 7.26 & 10 & 12.28 & 7.75 & 79 & 89.48 & 52.85 & 52 & 60.47 & 43.22\\
N=6 & 7 & 8.31 & 5.86 & 8 & 9.47 & 6.59 & 55 & 64.05 & 42.43 & 34 & 43.26 & 35.29 \\
N=8 & 5 & 6.18 & 4.73 & 6 & 7.4 & 5.57 & 39 & 47.27 & 34.99 & 23 & 31.6 & 28.98\\
\bottomrule
\end{tabular}
\end{adjustbox}
\label{tab:comp}
\end{table*}

            

\subsection{Classifying locations based on frequency and attractiveness}

We conduct a first experiment to classify the location based on the taxonomy  presented in Section 4.3; next we highlight an interesting property of 'significant' locations in summary trajectories. 

\noindent
Given a dataset $D$, for every  trajectory $T_i\in D$, and $\widehat{T_i} \in \widehat{D}$, we determine the classes of (distinct) locations labeled \emph{significant, transit, sporadic, insignificant} in accordance with the definition given in Section~\ref{sec:location_taxonomy}. Let us denote the generic class as  $X_i \in \{SL_i, TL_i, PL_i, IL_i\}$, respectively. At population scale, the percentage of locations in class $X$ is simply computed as average:  $$X\%=\frac{1}{|D|}\sum_{i=1}^{|D|} \frac{|X_i|}{|T_{u_i}|}$$


\noindent
Table~\ref{tab:location_type_percentage} reports 
the percentages of locations in each class computed for the four datasets of summary trajectories (i.e., $N=2,4,6,8, \delta=16'$) . 
It can be seen that:
\begin{itemize}
\item 
 The vast majority of locations in native trajectories are classified as \textit{insignificant}. The percentage varies from approx. 67\% to 84\%. Thus, the two approaches are in agreement that around  70\% of the locations in  the native trajectory are not important for the user. This result is substantially in line with the literature.
 \item  The percentage of \emph{significant} locations, i.e.  frequent and attractive, computed for the whole set of locations visited by an individual varies between 9\% and 19\%.
 
\item Transit (frequent but not attractive) and sporadic (infrequent but attractive) locations, together,  account for  7-14\%, on average, of the locations. 
\end{itemize}

\begin{table}[h]
\caption{Percentage of locations in native trajectories classified accordingly to the proposed taxonomy}
\centering
\label{tab:location_type_percentage}
{

\begin{tabular}{l c c c c}
\toprule
Location class     & \multicolumn{1}{c}{n=2} & \multicolumn{1}{c}{n=4} & \multicolumn{1}{c}{n=6} & \multicolumn{1}{c}{n=8} \\ \midrule
Significant (SL)   & 18.67 \%                 & 14.37 \%                 & 10.96 \%                  & 8.61 \%                  \\ 
Transit (TL)       & 7.13 \%                  & 5.74 \%                  & 4.43 \%                  & 3.41 \%                  \\ 
Sporadic (PL)      & 7.13 \%                  & 5.74 \%                  & 4.43 \%                  & 3.41 \%                  \\ 
Insignificant (IL) & 67.05 \%                 & 74.16 \%                 & 80.17 \%                 & 84.58 \%                 \\ \bottomrule
\end{tabular}

}
\end{table}

\begin{figure*}[t]
\centering
\centering

 \includegraphics[width=.8\textwidth]{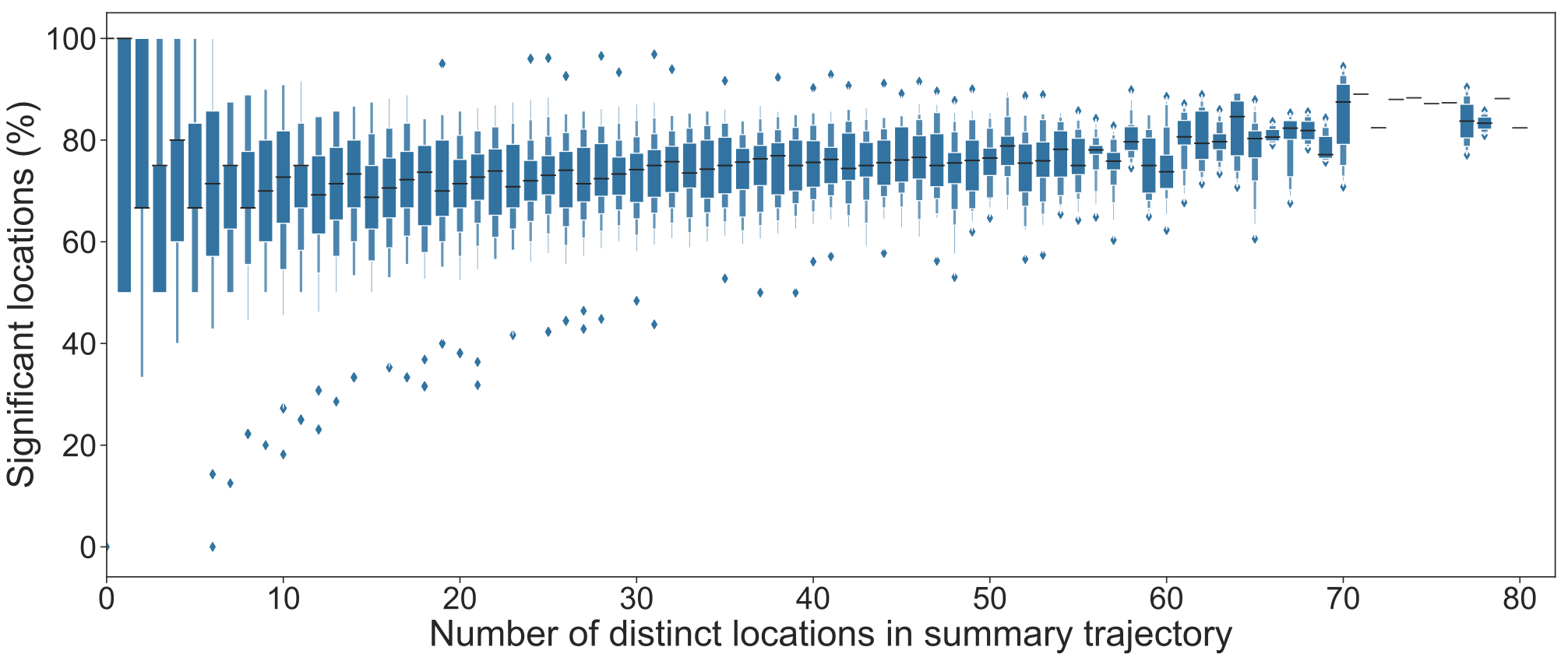}
\caption{Boxplot of the percentage of significant locations in the summary trajectories with parameters set $N = 2$, $\delta = 16'$. Each bar represents the distribution of the percentage of significant locations of all the trajectories of a given number of distinct attractive locations (the value on the x-axis)}
\label{fig:perc_common}
\end{figure*}

An interesting aspect to investigate is the possible dependence of the percentage of significant locations on the number of attractive locations.
Thus, we zoom in on summary trajectories, i.e. for every summary trajectory, we compute the ratio between the cardinality of the set of significant locations and the number of distinct locations in the summary trajectory.
%
The boxplot in \figurename ~\ref{fig:perc_common} reports the percentage of significant locations in the summary trajectories, grouped by the number of distinct locations, i.e. each bar represents the distribution of the percentage of significant locations of all the trajectories with a given number of distinct locations (the value on the x-axis). 
It turns out that the majority of attractive locations, around $70\%$, are the most frequently visited, too. Surprisingly, we find that this result does not depend on the number of distinct locations in the summary trajectory, i.e. it  holds  both for people  with high mobility and for more settled people visiting a limited number of locations.

\noindent

\noindent

\subsection{Utility of summary trajectories}
The question is whether summary trajectories preserve key statistical properties of human mobility data, that are found in literature, 
and, thus, data utility.
We perform two experiments:
(a) the first experiment is to analyze the location rank distribution in native and summary trajectories;
(b) the second one is to evaluate the level of matching between the two location rankings.

\subsubsection{Analysis of the location rank distribution}

\noindent

For this experiment, we refer to a key result in literature, that the frequency of visits exhibits a heavy-tail distribution, where few locations account for more than half of visiting frequencies, a limited set of locations are visited occasionally, while the long tail accounts for a large number of locations visited rarely or even once \cite{barabasi2008}.
To show that this mobility dataset follows the same behavior and that the summarization algorithm does not alter the nature of this primary feature of human mobility, 
we follow the approach in \cite{barabasi2008}, in particular, we compute the probability distribution of the location rank over the user population.

\noindent

\noindent
Given a generic telco dataset $D$, we define the probability $P(r,D)$ to find a user in a location with a  rank $r$ as follows. 
For every trajectory $T \in D$, we compute the visiting frequency of every location $l \in T$ and assign $l$ a univocal rank $r$ in the range $[1, |L|]$. Hence, the probability $P(r, T)$ to find the user in the location of rank $r$, is given by the number of occurrences of the location of rank $r$,  $occ(r)$, over the length of the trajectory, denoted:
 $P(r, T)=\frac{occ(r)}{|T|}$.
Scaling up, we define the average probability $P(r, D)$ to find a user in a location of rank $r$ as: 
\begin{equation}
\label{ff}
P(r, D)=\frac{1}{|D|}\sum_{T_i \in D} {P(r, T_i)}
\end{equation}

If we compute this measure of probability over the sets $D$ and $\widehat{D}$, i.e. $P(r, D)$, $P(r, \widehat{D})$, we obtain two distributions of the location rank that can be compared.

\begin{figure}[t]
    \centering
    \includegraphics[width=0.7\linewidth]{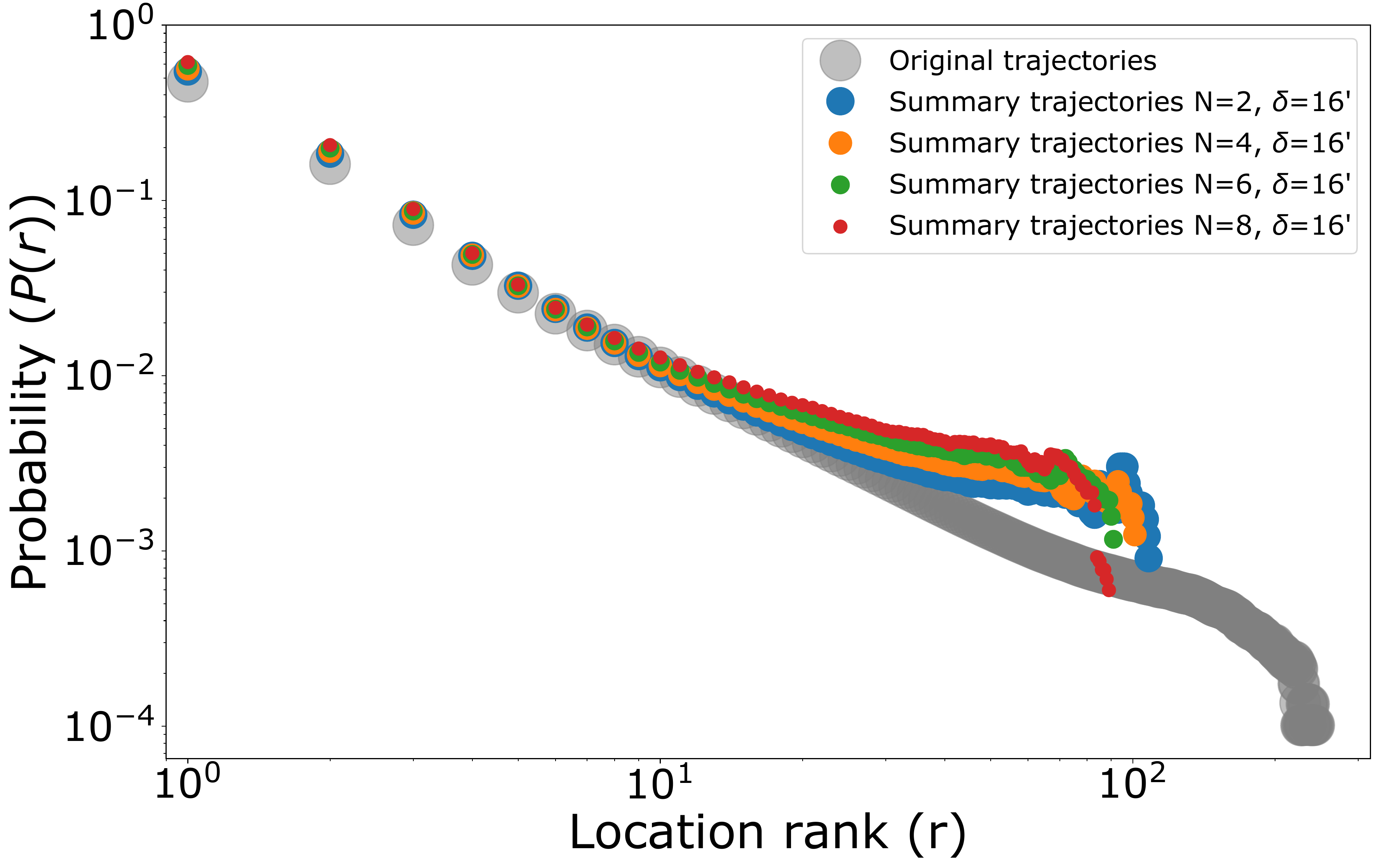}
    \caption{Plot in log-log scale of the average probability to find users in their locations as ranked by the frequency of visit. x-axis: locations ordered by descending frequency of visit and labeled by their rank. y-axis: probability to find  users at a ranked location computed as the average percentage of times users were recorded in that location}
    \label{fig:barabasi}
\end{figure}

\noindent
The log-log plot of these two distributions is reported in Figure \ref{fig:barabasi} for multiple choices of the two parameters of the summarization algorithm.
Referring to the red curve, which provides the most different trajectory from the original one, we find that the average probability for users to be in their first relevant locations is very high and equal to 0.48 and 0.61 for $r=1$, to 0.16 and 0.21 for $r=2$, and to 0.07 and 0.09 for $r=3$ in the original and summary trajectories, respectively. We find that these values are reassuringly aligned with the literature.
The discrepancies are even smaller for the other parameter settings, i.e. with $N$ decreasing, highlighting the consistency between the performance measures and the location ranking frequency. In fact, as observed in Figure \ref{AvgQ_S_many_params}, the lower the value of $N$, the higher the goodness of the summarization. In particular, with $N=2$ we obtain the best summarization goodness in Figure \ref{AvgQ_S_many_params} and the closest curve to the original trajectories one in Figure \ref{fig:barabasi} \cite{barabasi2008}.
Figure \ref{fig:barabasi} also shows that the overall shape of the curve is not altered and the visiting probabilities are very similar in the original and summary trajectories up to around the first most visited ten locations. The summarization mainly impacts the long right tail which is shortened as the low-ranked locations are discarded by the clustering algorithm. This first result only accounts for the fact that the algorithm does not alter the probability of visiting the locations for high ranks, while eliminating the tail of visiting probabilities of low ranking locations. However, this result does not provide any evidence on the matching between the locations and their rank in the original and summarized trajectories. We can here conclude that, \emph{the summarization technique is able to preserve the frequency of the highly ranked, while discarding the locations with very low ranks.}
To fully affirm that the summarization technique really preserves the locations regularly frequented, while discarding the locations that are largely irrelevant to a person as they have only been there even once or may be passed through them once by chance, we will investigate the degree of  matching in the following section.

\subsubsection {Analysis of the matching degree}

The problem is formulated as follows: for each trajectory, determine the maximum value $k$, such that the top-k most frequented locations in $T_u$, denoted $N_u(k)$, are also attractive, i.e. included in  $\widehat{T_u}$. 



\noindent

\begin{figure*}[t]
    \begin{subfigure}[]{.485\linewidth}
        \includegraphics[width=\textwidth]{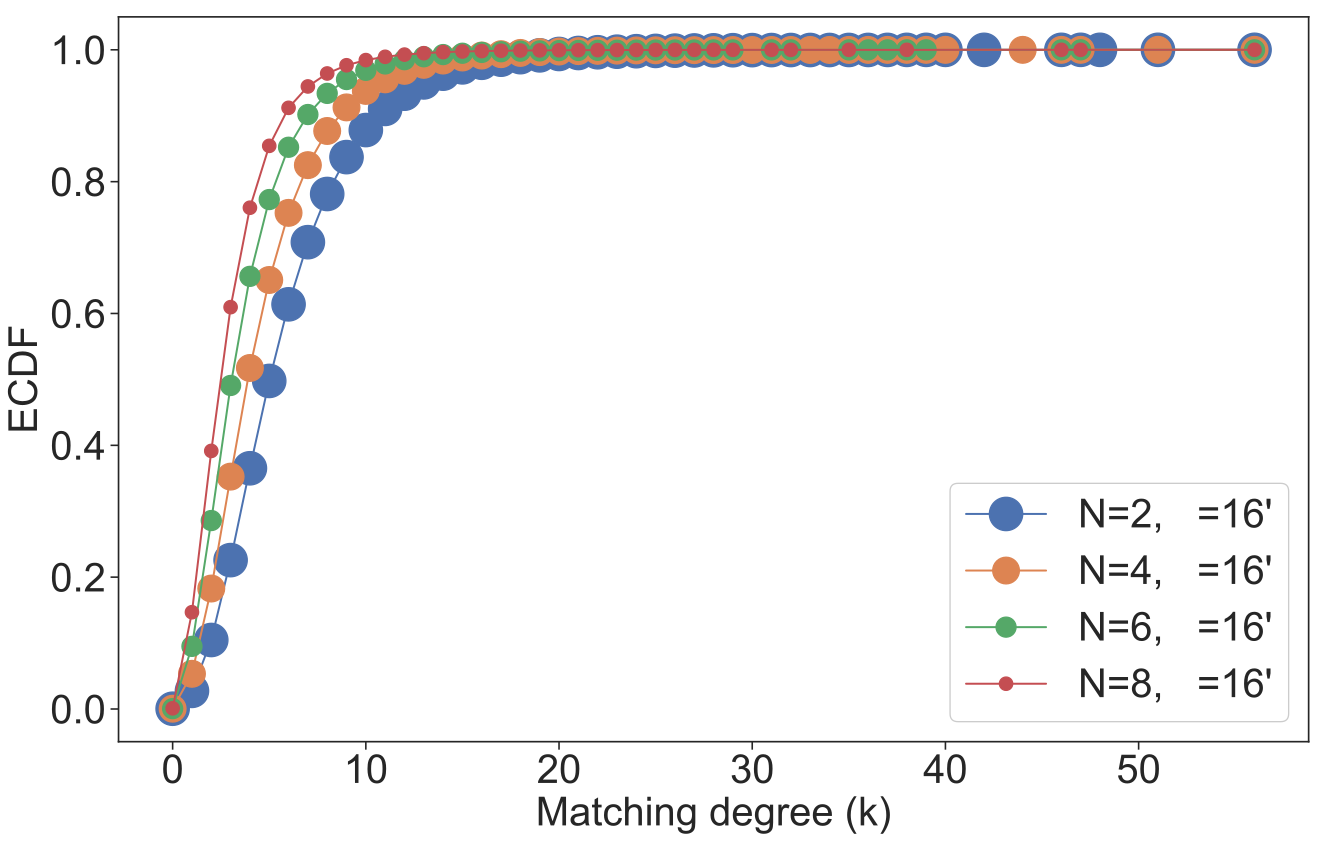}
    \end{subfigure}
    \begin{subfigure}[c]{.485\linewidth}
        \begin{tabular}{c c c c}
            \toprule
             Parameters & Median &  Mean &  Std \\
             \midrule
             N=2, $\delta$=16$'$ & 6.0 &  6.31 & 3.77  \\
             N=4, $\delta$=16$'$ & 4.0 &  5.10 & 3.20\\
             N=6, $\delta$=16$'$ & 4.0 &  4.16 & 2.71\\
             N=8, $\delta$=16$'$ & 3.0 &  3.49 & 2.32\\
             \bottomrule
        \end{tabular}
    \end{subfigure}
    \caption{Empirical Cumulative Distribution Function (ECDF) and summary statistics of the matching degree $k$}
    \label{fig:cdf_k}
\end{figure*}

\figurename ~\ref{fig:cdf_k} reports the cumulative distribution function of the variable $k$ for different values of the parameters of the summarization algorithm.
%
The value of $k$ is around 5, meaning that the summarization algorithm captures the 5 most frequently visited locations, on average. Therefore, importantly,  the summarization can detect not only the routine locations as home, work and similar, but also locations which are out of the daily routine  and only occasionally visited. 

\begin{figure}[h]
    \centering
        \includegraphics[width=0.8\textwidth]{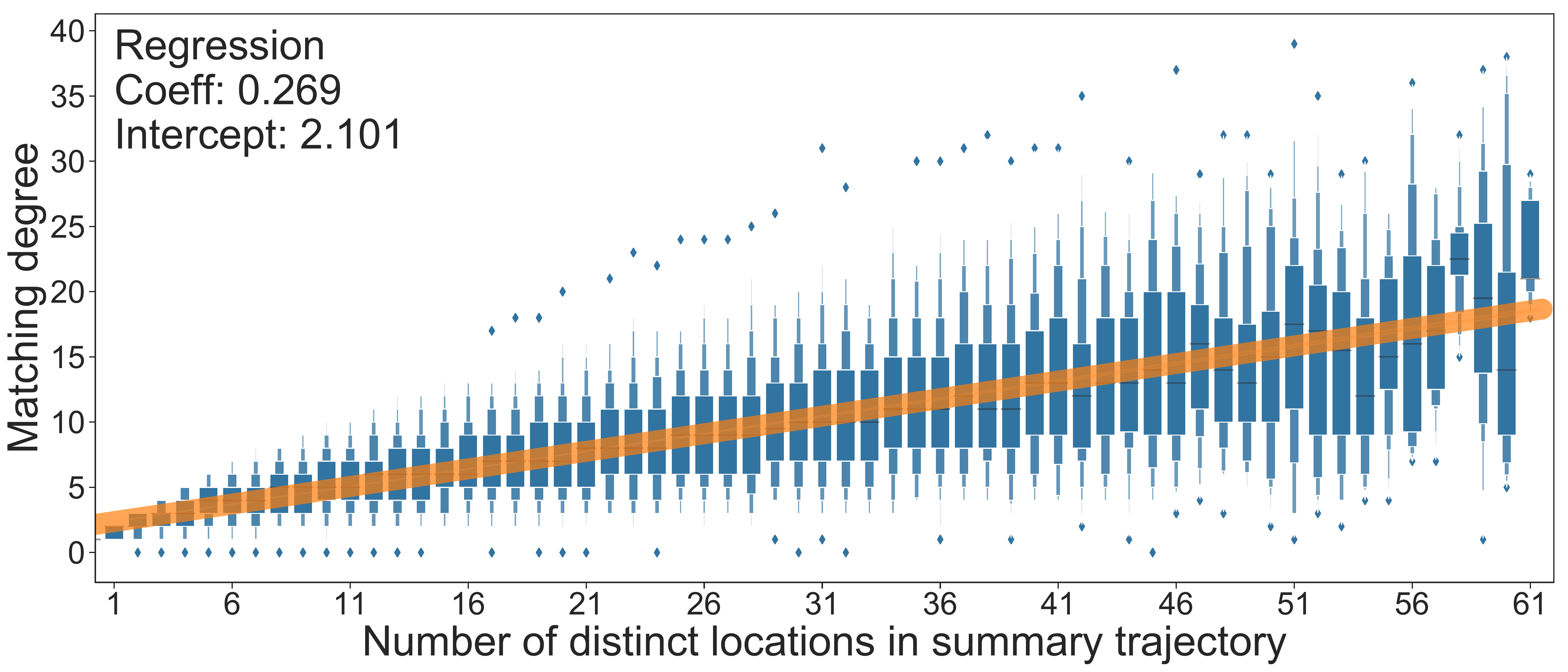}
    \caption{Boxplot of  the matching degree $k$ as a function of the number of distinct locations in the summary trajectory with parameters set $N = 2$, $\delta = 16'$. Each bar represents the distribution of the matching degree of all the trajectories with a given number of distinct attractive locations (the value on the x-axis). The orange line represents the linear regression of the matching degree over the number of attractive locations}
    \label{fig:regr}
\end{figure}

\noindent
A related question is whether the matching degree depends on the length of the trajectory. \figurename ~\ref{fig:regr} plots the matching degree as function of the number of distinct locations in the summary trajectory. Interestingly, it can be seen that the matching degree $k$ tends to increase linearly with the number of attractive locations. 

\subsection{Evaluating location diversity}
We perform two experiments:
(a) the analysis of location diversity profiles based on summary trajectories;
(b) the analysis of the entropy rate.

\subsubsection{Analysis of location diversity profiles} This experiment is organized in two steps: the first step is to compute the location diversity profile of every single trajectory; the second step is to categorize users based on location diversity indices.


\begin{table}[h]
	\centering
	\caption{Summary statistics on the variables of the location diversity profiles.}	
	\begin{tabular}{c c c c c}
		\toprule
		Metric & Max  & Min & Avg   & Std  \\
		\midrule
		$R$ & 108 & 1 & 15.61 & 8.87 \\
		$TD_H$ & 91 & 1 & 6.95 & 3.84\\
		$TD_S$ & 74.6 & 1 & 4.49 & 2.41\\
		\bottomrule	
	\end{tabular}	
	\label{tab:statsmetrics}
\end{table}
\noindent
We consider the summarized dataset $\widehat{D}$ obtained by setting the parameters: $N=2$, $\delta=16'$.
For every trajectory  $\widehat{T_i} \in \widehat{D}$, we compute the diversity profile:  $Pr(i)=(R^i, TD_H^i, TD_S^i)$. The summary statistics for the variables of the profile are reported in Table \ref{tab:statsmetrics}. It can be seen that, on average, a trajectory contains about  15 relevant  locations, while the true diversity of order 1 ($TD_H$)  amounts to about 7 locations, and the true diversity of order 2 ($TD_S$) to about 4 locations. There is thus a significant gap between the values of $R$ and $TD_x$, indicating that locations are not evenly frequented. Moreover, the diversity of core locations is quite small. Further details are provided in the next step.   

\noindent
%
User classification is performed as follows: for every variable (i.e., $R, TD_H, TD_S$), we consider the sorted sequence of values (one value per trajectory). Every sequence is next partitioned in a series of 1-D clusters using the Jenks algorithm \cite{jenks1967}, finally, the relative size of every cluster is  plotted as a histogram.

\begin{figure}[h] 
	\centering
	\subcaptionbox{Classification based on $R$} 
	{\includegraphics[width=.45\textwidth]{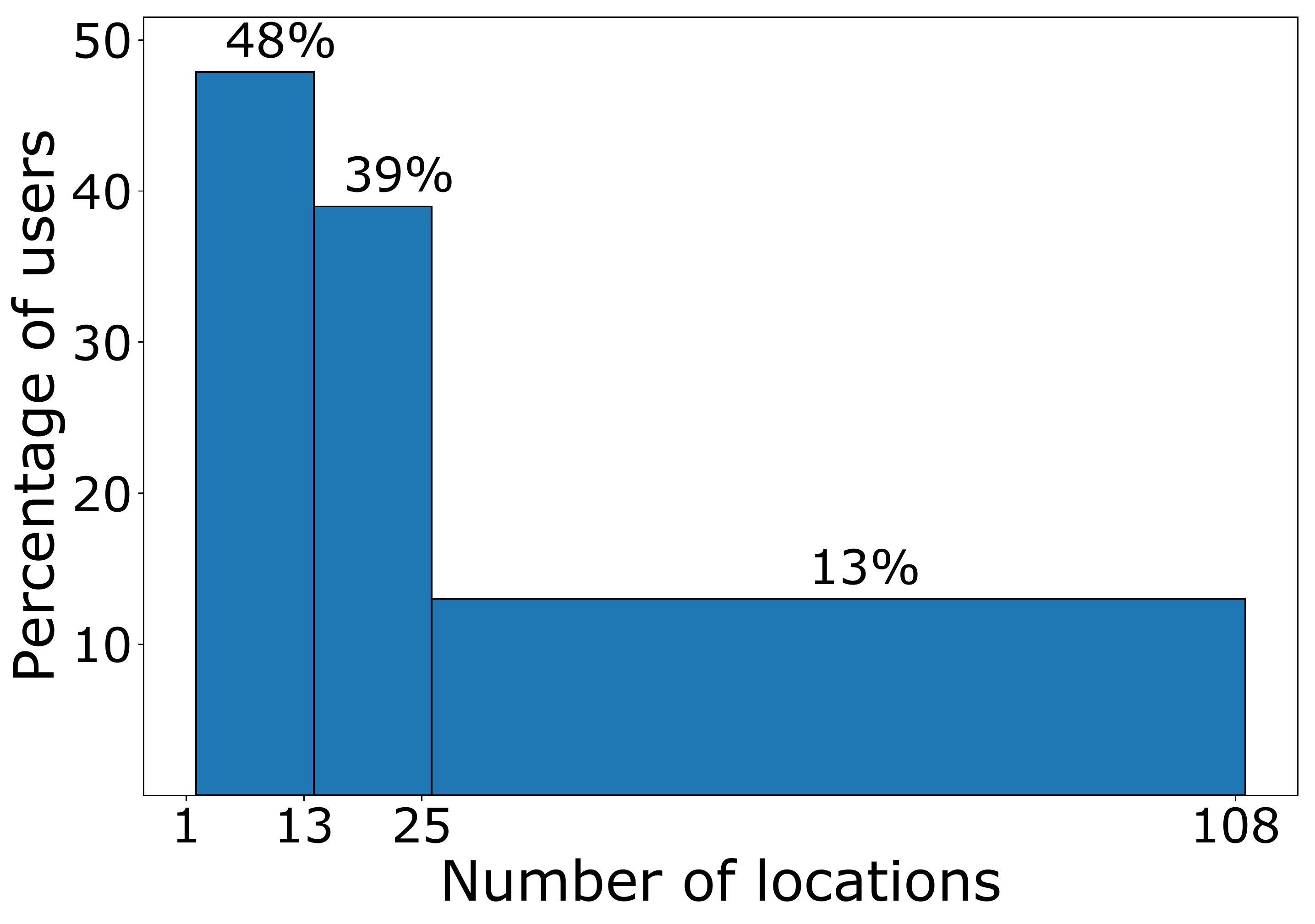}}\\
	\subcaptionbox{Classification based on $TD_H$}
	{\includegraphics[width=.45\textwidth]{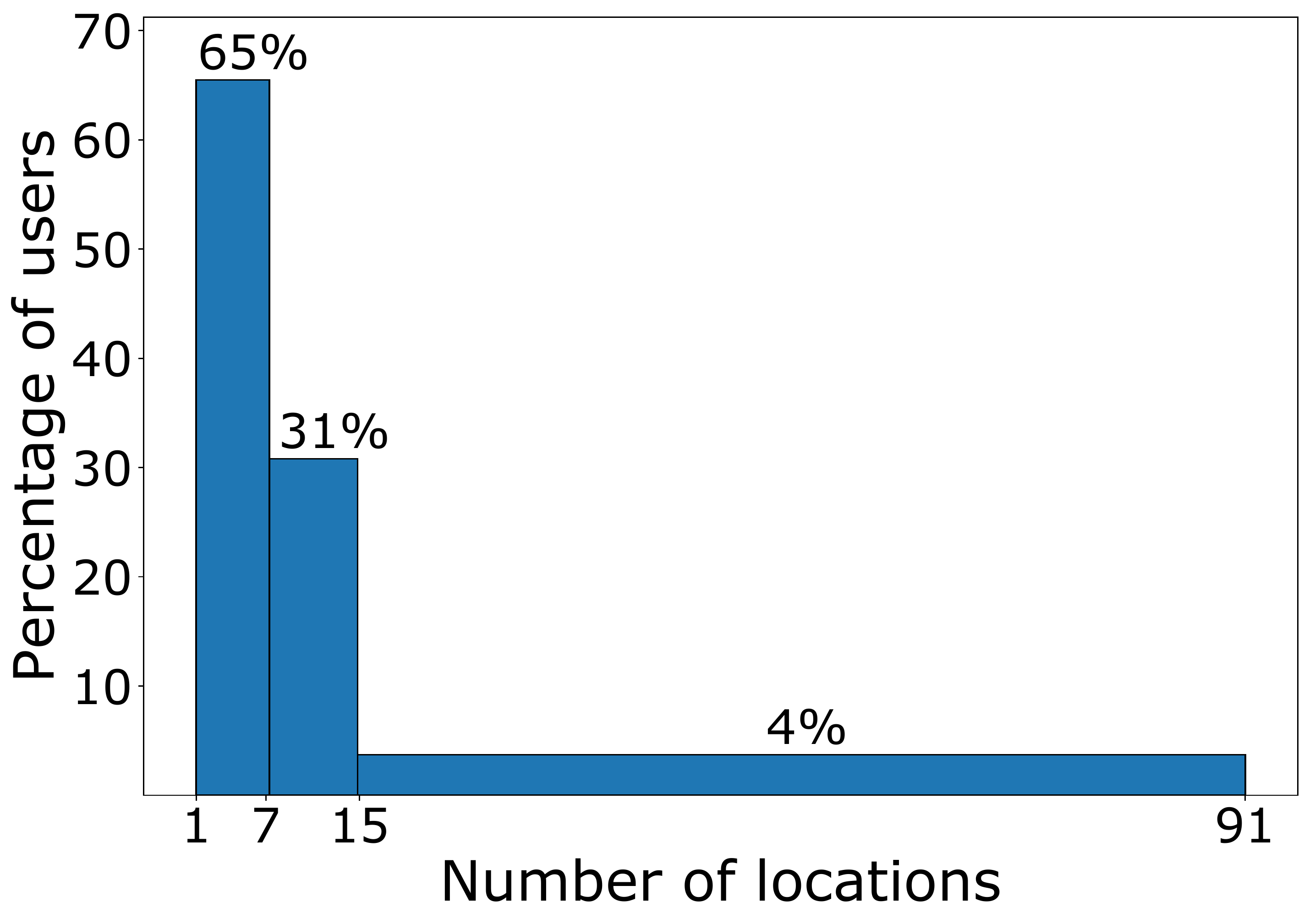}}
\subcaptionbox{Classification based on $TD_S$}
	{\includegraphics[width=.45\textwidth]{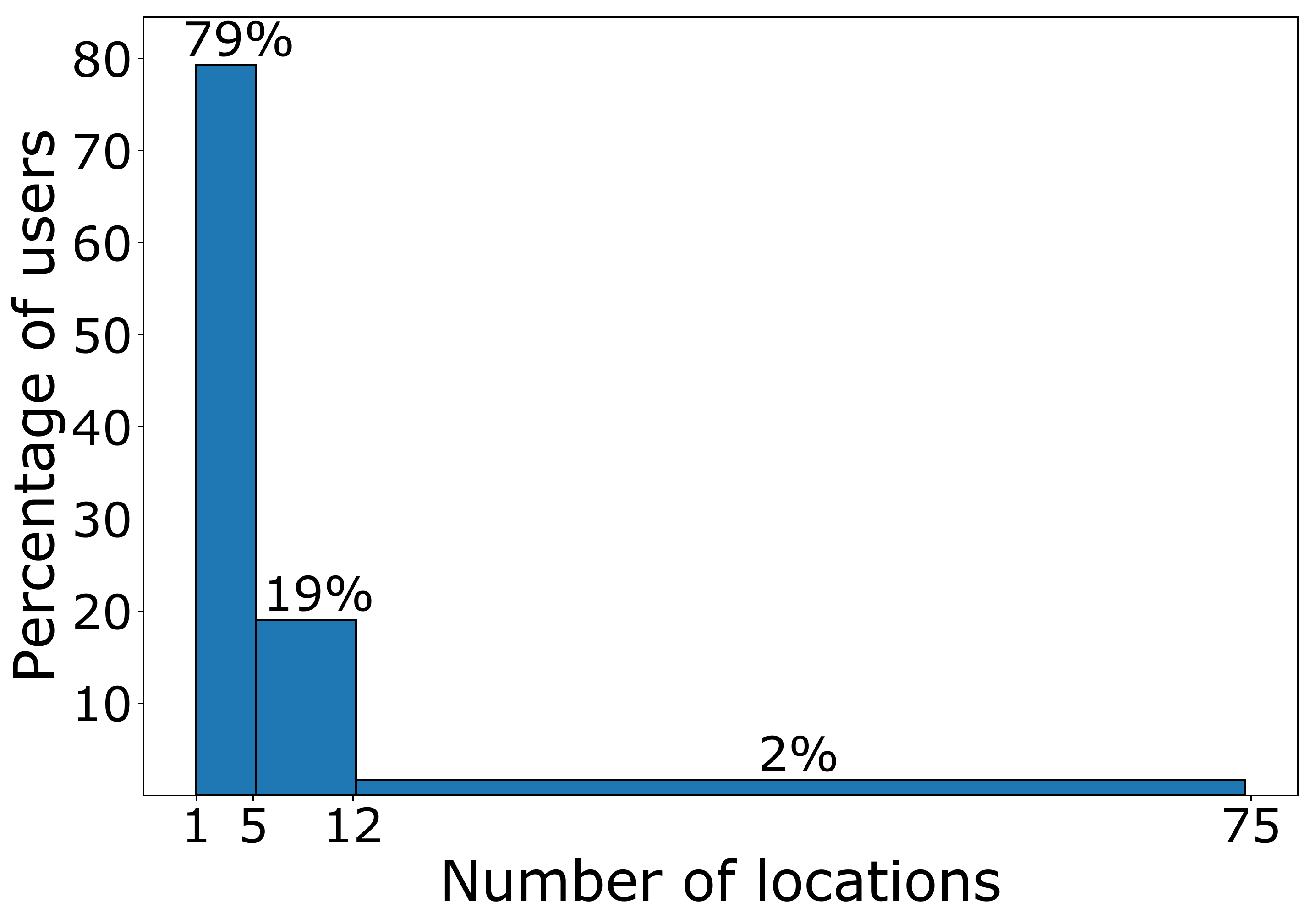}}
	\caption{User classification based on diversity indicators. The x-axis reports the number of locations, the y-axis the percentage of users} 
	\label{fig:jenks}
\end{figure}
The Jenks algorithm is a data classification method that optimizes the arrangement in classes of 1D values by minimizing the within-class
variance and maximizing between-class variance. As a result,  the elements within each class are as close as possible in value to each other. 
In more detail, the algorithm splits a sequence of sorted data (unidimensional) into $k$ intervals so as to minimize the  internal  measure of quality, the \textit{goodness-of-variance-fit (GVF)} \cite{jiang2013}. GVF ranges between 0 and 1, with 0 indicating the worst fit and 1 the best fit. To determine a suitable  value for $k$, the empirical rule is to run the algorithm for increasing value of $k$ until the GFV exceeds a threshold value. With a threshold of 0.7, we obtain $k=3$.  

The plots in Figure \ref{fig:jenks} report, for each diversity measure, the three classes, i.e. low, middle and high  diversity, along with the percentage of users falling in each class. 
Figure \ref{fig:jenks}(a)  shows the classification based on $R$. It can be seen that for 87\% of users, the number of relevant locations is at maximum 25, while  for 13\% of users, it ranges between 25 and 108 locations,
Figure \ref{fig:jenks}(b)  shows a refined classification based on $TD_H$:  for 65\%  of users, the diversity of ``typical location''  amounts to 7 locations, while only for 4\% of users the diversity is greater than 15 locations. 
Finally, Figure \ref{fig:jenks}(c) shows the further refined classification based on $TD_S$. We see that, for 79\% of users, the maximum diversity is 5 locations, while only for 2\% of users, the diversity is greater than about 12 locations.

These results are qualitatively in line with what is reported in literature (see Section \ref{rw}). We find particularly interesting the inverse of the Simpson Index, i.e. $TD_S$, as it provides a measure of the diversity of core locations,  typically home, work, a few others. Such a measure is in line with what is known, that  the great majority of people frequent assiduously only few locations.

\subsubsection{Entropy rate at population scale}

\begin{figure}[h]
    \centering
    \includegraphics[width=0.67\textwidth]{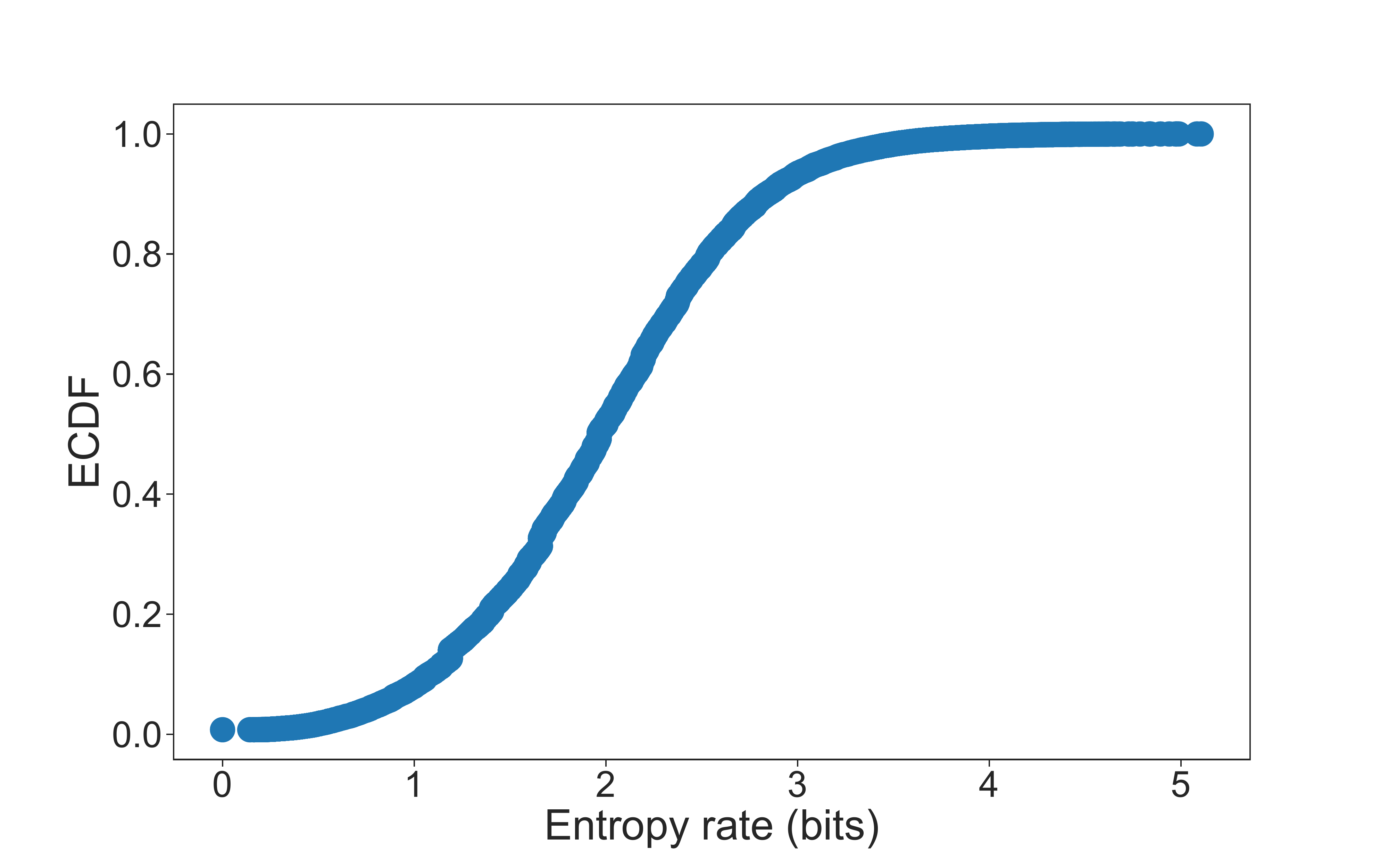}
    \caption{Empirical Cumulative Distribution Function (ECDF) of the entropy rate of summary trajectories of the reference dataset}
    \label{fig: entropy rate CDF}
\end{figure}

We have estimated the entropy rate of summary trajectories of the reference dataset. Summary trajectories have a rather short length (typically a few dozen symbols) and we found that the worst-case $O(n^3)$ complexity of Algorithm~\ref{alg: entropy rate estimation} is not a limitation in practice, being able to estimate the entropy rate of a typical summary trajectory in about $10\,\textrm{\textmu s}$ on a standard PC.
The cumulative distribution function of the entropy rate of summary trajectories of the reference dataset is shown in Figure~\ref{fig: entropy rate CDF}.
The mean value is 1.97 bits; the most complex trajectory has an entropy rate of 5.11 bits.

\section{Discussion}
In this work, we use a research methodology that combines computational models with methods  taken from human mobility modeling and ecology, for  analyzing the nature of the locations frequented by individuals and thus their mobility behavior. The main findings of this effort can be summarized as follows: 
   \begin{itemize}
    \item 
    We have introduced the notion of trajectory summarization and  presented the SeqScan-d algorithm for the extraction of \emph{attractive} locations from symbolic trajectories with repeating symbols, in particular telco trajectories. 
    SeqScan-d 
    is capable of 
    identifying key locations for each user and dispensing with the irrelevant ones, despite the high noise and uncertainty of the data. Unlike methods relying on the compression of symbolic sequences, for example based on RLE, this technique is robust against noise. That is important for the effectiveness of summarization, in that it reduces fragmentation, while preserving the semantic coherence of location visits. 
%
    Concerning the parameters,  the summarization goodness index is optimal with N=2. 
    Consequently, solely the parameter $\delta$ needs to be specified, based on the applications characteristics. That streamlines the use of the operation.   
 With the parameters $N=2$, $\delta=16'$, the number of locations (as types) is reduced by 74\% on average. 

 
 \item We have devised a framework for classifying locations along  the two dimensions of frequency and attractiveness. The result is a novel location taxonomy which consists of four formally defined location classes. The class of insignificant locations accounts for $67\%$ of the locations in the native trajectory with a very high concordance between the frequency-based and the attractive-based approach. The taxonomy highlights two particularly interesting classes, the \emph{transit} and the \emph{sporadic} locations, that on average accounts for the $14\%$ of the locations with the parameters $N=2$, $\delta=16'$. Transit locations are for example the areas crossed by a traveler.  Therefore, if a trajectory presents a high percentage of transit locations, the user is likely very mobile. Sporadic are  the locations that, though important for the user, are only occasionally visited.  A trajectory with many sporadic locations is typical of an individual with a non-routinary life. 
 In synthesis, we show that the two relevance models -  frequency-based and attractiveness-based - are not in contrast, but rather offer complementary viewpoints on the user behavior. 
 
\item  We have shown  that summary trajectories preserve the basic and universal statistical property of human mobility, that the frequency of visits follows a heavy-tail distribution, where few locations account for more than half of visiting frequencies, a limited set of locations are visited occasionally, while the long tail accounts for a large number of locations visited rarely or even once. Additionally, it is shown that the summarization operation captures the 6 most frequently visited locations, on average, with the parameters $N=2$, $\delta=16'$ . 
\item 
   We have introduced the  notion of location diversity profile to capture  different and multi-level facets of location diversity. This is another way for characterizing the individual mobility, this time at trajectory level and not at location level, as above. 
    We have seen, for example, 
    that the diversity of the locations based on the Shannon-Weiner measure amounts, on average, to seven locations (based on the exponential of the Shannon-Weiner measure), while the diversity  based on the inverse Simpson index  to about four locations. These results, though computed on summary trajectories, are  in line with the state-of-the-art. Finally, we have introduced the  entropy rate to measure entropy over sequences. Although the use of this metric for the modelling of human mobility is known, we believe it can be of broad interest also for the spatial computing community.  
\end{itemize}

\section{Conclusion}
To summarize, in this article we have presented a location-centric framework for the analysis of the individual mobility behavior. Such analysis relies on symbolic  trajectories resulting from a process of 'semantic compression' of telco data. The code of SeqScan-d is publicly available (https://github.com/SeqScan/SeqScan-D).  
 We have also emphasized the distinction between places of attractiveness, where users are willing to spend their time, and the locations that are frequent but not relevant  where users often pass but without consuming  time as those locations are not attractive to them. This distinction can be of interest 
in all those applications that focus on places of interest for people such as  epidemics, where the identification of places where people spend time is fundamental, or recommendation systems focused on the attractiveness of places. 
The third and last core topic -  perhaps the most challenging - regards the specification of suitable metrics for measuring the mobility behavior of individuals. 
This is a complex topic that requires further investigation along diverse directions, e.g. \cite{2020Damiani}. Measuring the mobility behavior can be of practical interest in many applications in which the movement of people can be naturally represented in symbolic form, such as indoor movement analysis, e.g., measuring the movement in shopping areas, museums, airports, or in urban settings. 
To conclude, the study of human mobility calls for a strong multidisciplinary effort to integrate data management and related application areas, e.g. urban computing, with mathematically grounded human mobility  models. 


\section*{Acknowledgments}
The research carried out by Maria Luisa Damiani,  Fatme Hachem and Matteo Rossini is partially supported by the Italian
government via the NG-UWB project (MIUR PRIN 2017). We thank the reviewers for the insightful comments.

\bibliographystyle{unsrt}  
  \bibliography{references.bib}

\end{document}